\documentclass[12pt]{article}
\usepackage{amsmath}
\usepackage{graphicx,psfrag,epsf}
\usepackage{enumerate}
\usepackage{url} % not crucial - just used below for the URL 

\newcommand{\blind}{0}

\usepackage[table,dvipsnames]{xcolor}
\usepackage{amsthm}
\usepackage{mathtools}
\usepackage{collcell}
\usepackage{dsfont}
\usepackage{rotating}
\usepackage{wrapfig}
\usepackage[utf8]{inputenc}
\usepackage{tikz}
\usepackage{lettrine}
\usepackage{dblfloatfix}
\usepackage[english]{babel} 
\usepackage{amsfonts}
\usepackage{subfig}
\graphicspath{ {fig/} }
\usepackage[hyperfootnotes=false]{hyperref}

\usepackage{lipsum}
\usetikzlibrary{arrows,decorations.markings}
\usepackage{standalone}
\usepackage[numbers, sort&compress]{natbib}
\usepackage{changes}

\definecolor{gray}{rgb}{0.5,0.5,0.5}
\definecolor{red}{rgb}{0.8,0,0}
\definecolor{dred}{rgb}{0.5,0,0}
\definecolor{blue}{rgb}{0,0.1,1}
\definecolor{dblue}{rgb}{0,0.1,0.6}
\definecolor{cyan}{rgb}{0,0.5,.5}
\definecolor{dcyan}{rgb}{0,0.3,.3}

\begin{document}
	\setcitestyle{square,sort&compress}

	\def\spacingset#1{\renewcommand{\baselinestretch}%
		{#1}\small\normalsize} \spacingset{1}

	%%%%%%%%%%%%%%%%%%%%%%%%%%%%%%%%%%%%%%%%%%%%%%%%%%%%%%%%%%%%%%%%%%%%%%%%%%%%%%
	
	\if0\blind
	{
		\title{\bf Ensemble Forecasting for Intraday Electricity Prices: Simulating Trajectories\footnote{accepted for publication in Applied Energy.}}
		\author{Michał Narajewski\hspace{.2cm}\\
			University of Duisburg-Essen\\
			and \\
			Florian Ziel \\
			University of Duisburg-Essen}
		\maketitle
	} \fi
	
	\if1\blind
	{
		\bigskip
		\bigskip
		\bigskip
		\begin{center}
			{\LARGE\bf Title}
		\end{center}
		\medskip
	} \fi
	
	\bigskip
	\begin{abstract}
Recent studies concerning the point electricity price forecasting have shown evidence that the hourly German Intraday Continuous Market is weak-form efficient. Therefore, we take a novel, advanced approach to the problem. A probabilistic forecasting of the hourly intraday electricity prices is performed by simulating trajectories in every trading window to receive a realistic ensemble to allow for more efficient intraday trading and redispatch. A generalized additive model is fitted to the price differences with the assumption that they follow a zero-inflated distribution, precisely a mixture of the Dirac and the Student's t-distributions. Moreover, the mixing term is estimated using a high-dimensional logistic regression with lasso penalty. We model the expected value and volatility of the series using i.a. autoregressive and no-trade effects or load, wind and solar generation forecasts and accounting for the non-linearities in e.g. time to maturity. Both the in-sample characteristics and forecasting performance are analysed using a rolling window forecasting study. Multiple versions of the model are compared to several benchmark models and evaluated using probabilistic forecasting measures and significance tests. The study aims to forecast the price distribution in the German Intraday Continuous Market in the last 3 hours of trading, but the approach allows for application to other continuous markets, especially in Europe. The results prove superiority of the mixture model over the benchmarks gaining the most from the modelling of the volatility. They also indicate that the introduction of XBID reduced the market volatility. 
	\end{abstract}
	
	\noindent%
	\vfill
	
	\newpage
	\spacingset{1.45} % DON'T change the spacing!
	
	%----------------------------------------------------------------------------------------
	%	ARTICLE CONTENTS
	%----------------------------------------------------------------------------------------
	
	\section{Introduction}
	
	Intraday continuous electricity markets gain on importance every day \cite{goodarzi2019impact}. Their primary purpose is to handle the uncertainty in electricity generation and load arisen since the day-ahead markets \cite{kath2018value}. A number of events can cause the uncertainty, e.g. unexpected power plant outage or changing weather conditions. The latter one is the result of the global trend of investing in weather-dependent renewable power sources and is a subject of modelling and forecasting \cite{maciejowska2020assessing}. The need of intraday continuous trading is fulfilled by the power exchanges and transmission system operators (TSO) \cite{Viehmann2017}. They allow the market participants to trade the energy continuously up to 5 minutes before the delivery, e.g. in France or Germany, and to trade it cross-border, e.g. using the cross-border intraday (XBID) market \cite{kath2019modeling}. Even though there is a clear evidence of the importance of this kind of markets, the researchers do not investigate them in terms of forecasting as willingly as the day-ahead market.
	
	The day-ahead market is the main electricity spot market with a long history of research on electricity price forecasting \cite{weron2014electricity}. Recent studies on the electricity price forecasting (EPF) in day-ahead markets consider i.a. the probabilistic forecasting and forecasting combination. \citet{nowotarski2018recent} present a review of probabilistic EPF and \citet{muniain2020probabilistic} use it to simulate peak and off-peak prices. \citet{marcjasz2018selection} combine point forecasts achieved using different calibration windows while \citet{uniejewski2019importance} and \citet{serafin2019averaging} do it for probabilistic forecasts.
	A very big part of the recent EPF literature are also hybrid models \cite{yang2017electricity, wang2017multi,  zhang2020adaptive} and neural networks \cite{xiao2017research,bento2018bat, keles2016extended}. Also the market integration plays an important role in price formation in both day-ahead and intraday markets what is elaborated by \citet{lago2018forecasting} and \citet{kath2019modeling}.

	The role of the intraday markets in the balancing of electricity systems was emphasized and explained by \citet{ocker2017german} and \citet{koch2019short} on the basis of the German electricity market. They observed that the introduction of the intraday continuous market in Germany partially led to a substantial decrease in the demand for balancing energy while the wind and solar energy generation increased. \citet{karanfil2017role} clarify the reason for the spread between day-ahead and intraday prices in Denmark, while \citet{maciejowska2019day} forecast the price spread between the day-ahead and intraday markets based on the Polish and German data. The continuity of the intraday market has encouraged the researchers to investigate the transaction arrival process \cite{narajewski2019estimation}, bidding behaviour  \cite{kiesel2017econometric, graf2020modeling, rintamaki2020strategic} and optimal trading strategies 
	\cite{aid2016optimal, ayon2017aggregators, glas2020intraday}. The impact of fundamental regressors on the price formation in the intraday market was examined by \citet{pape2016fundamentals}, \citet{gurtler2018effect}, and \citet{kremer2019fundamental}. 
	
	The literature on the EPF in the intraday markets is not that broad as in the day-ahead markets or as the one regarding other aspects of the intraday markets. \citet{monteiro2016short} and \citet{andrade2017probabilistic} conducted the EPF for the Iberian intraday market, however it is not a continuous market, and thus their studies are more similar to these on day-ahead markets.
	\citet{uniejewski2019understanding}, \citet{narajewski2019econometric} and \citet{marcjasz2020beating} performed the EPF in the German Intraday Continuous Market, while \citet{oksuz2019neural} in the Turkish Intraday market. An outcome of the second one was an indication of the weak-form efficiency of the investigated market. This was partially confirmed by \citet{janke2019forecasting}, who forecasted the distribution of prices during the last three hours of trading and concluded that forecasting of the central quantiles yields marginal improvement to the naive benchmark. However, \citet{marcjasz2020beating} managed to outperform the most recent price by using an ensemble of it and a lasso-estimated model.
	
\begin{figure}[t!]
		
		\includegraphics[width=1\textwidth]{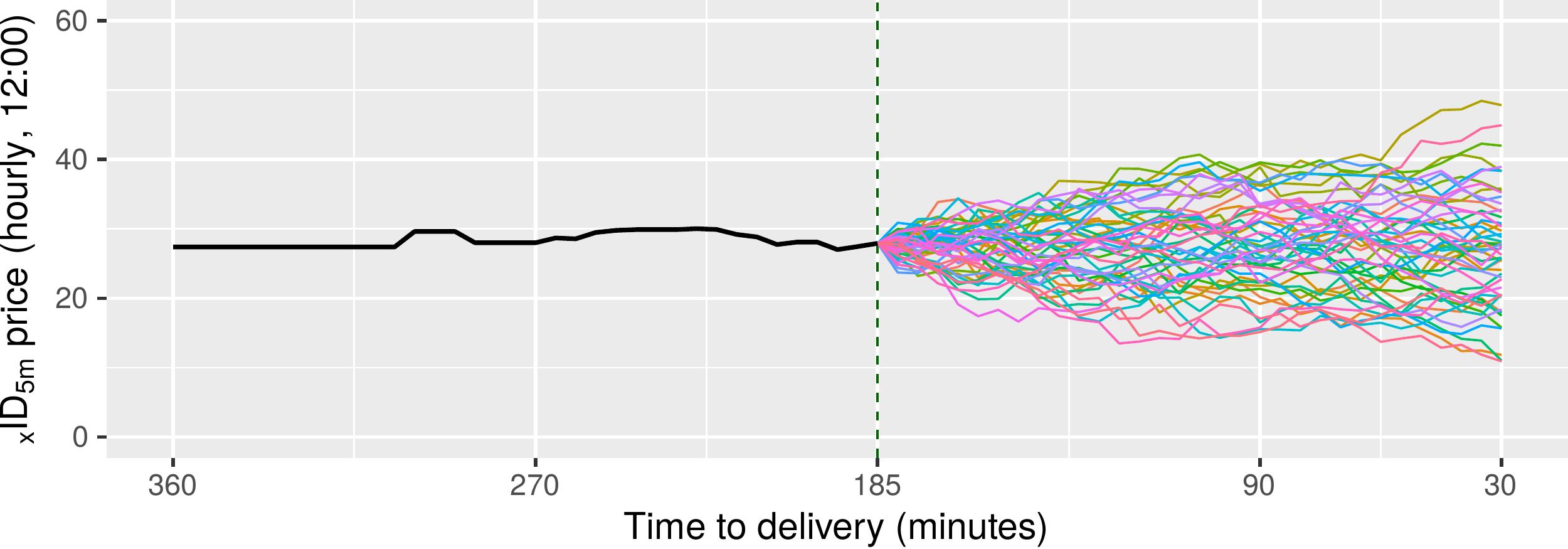}
		
		\caption{Price trajectory for the hourly product with delivery on 15.07.2016 at 12:00. The black part is the realization and the colourful part consists of 100 simulations from the Gaussian random walk. Time of forecasting is indicated by the green dashed line.}
		\label{fig:motivation}
		
	\end{figure}

	The only four papers on EPF in the German intraday market considered the ID$_3$-Price (a volume-weighted average price of transactions in the last three hours before delivery) as the most important price index in the German intraday market and conducted forecasting of it. This paper focuses on the ID$_3$ index as well, but not directly. Instead of forecasting its price we simulate the paths of 5-minute volume weighted average price during the time-frame of the index. This way we obtain a distribution forecast of the prices in every 5~min window during the last three hours before the delivery. An example of this approach can be seen in Figure \ref{fig:motivation}. We motivate our research with results on the weak-form efficiency of the market concluded by \citet{narajewski2019econometric} and a possible application of the methodology to trading of the electricity and optimal redispatch management.
	
	In purpose of modelling and forecasting of the trajectories, we utilize the generalized additive models for location scale and shape (GAMLSS) \cite{rigby2005generalized} which extends the generalized additive models (GAM) \cite{hastie1990generalized}. This methodology found applications to the electricity load \cite{pierrot2011short, gaillard2016additive} and day-ahead price \cite{serinaldi2011distributional, gianfreda2018stochastic, abramova2020forecasting} forecasting, but never to the intraday electricity markets. The model for price difference $\Delta P$ is fitted to the Student's t-distribution and mixed with the Dirac distribution, i.e. $\Delta P \sim (1-\alpha)\delta_0 + \alpha \text{t}$. $\alpha$ is assumed to be a Bernoulli variable with probability $\pi$ and is modelled using the logistic regression. We estimate it with the lasso method \cite{Tibshirani1996}. A broader description of the modelling exercise can be found in Section~\ref{sec:models}. 
	
	The forecasting part utilizes a rolling window study. This is a very common study type in the EPF and is widely utilized by researchers \cite{uniejewski2019understanding, narajewski2019econometric}. We analyse both in-sample characteristics and evaluate the out-of-sample forecasting performance. 
	
	The major contributions of this paper are as follows:
	\begin{itemize}
		\item[(1)] It is the first work on the price trajectories in intraday continuous markets which are a new and developing part of the electricity markets.
        \item[(2)] A rigourous presentation and discussion of all characteristics of the market, like trading frequency and volatility.
		\item[(3)] We propose a model that utilizes a mixture of GAMLSS and logit-lasso estimation methods and generates realistic ensembles what allows for efficient decision-making, especially for trading and redispatch.
		\item[(4)] The components of the proposed model are interpreted with respect to the market behaviour, highlighting the impact of the XBID introduction and relevant features, like wind and solar generation, load, calendar effects, trading activity and historic prices.
		\item[(5)] The high-quality predictive performance of the proposed model is compared with simple benchmarks and sophisticated models with respect to point and probabilistic forecasting.
	\end{itemize}
	
	The remainder of this paper has the following structure. In the next section, we describe the market. The third section consists of the data description and descriptive statistics. Then, a broader explanation of the estimation methods is presented, followed by the description of the considered models and benchmarks. In the fifth section, the forecasting study and evaluation measures are introduced and discussed in detail. In the sixth section, we present the results which consist of the in-sample analysis with relevant model interpretations and the out-of-sample evaluation.
	%, and the application example.
	The final section concludes this paper. The methodology used in the paper is very innovative, especially in regard to the intraday electricity markets. We present it with an application to the German Intraday Continuous Market, but it can be easily used with any other intraday electricity continuous market.

	\section{Market description}
	
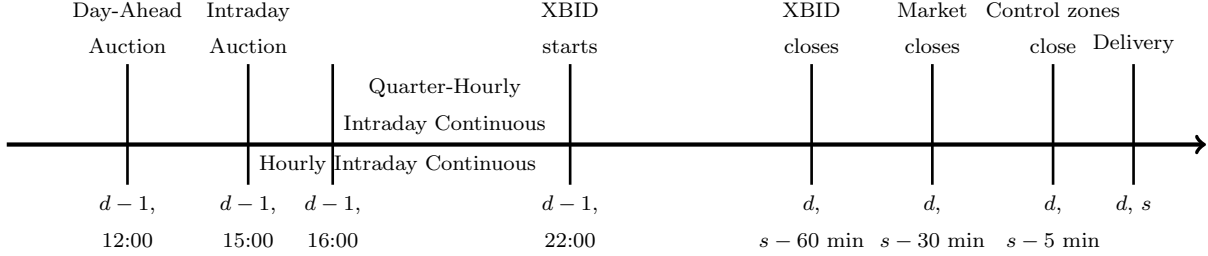
\begin{figure*}[!t]
		\begin{tikzpicture}[scale=1.07]
		\draw [->] [ultra thick] (0,0) -- (14.9,0);
		\draw [line width = 1] (1.5,1) -- (1.5, -0.5);
		\node [align = center, below, font = \scriptsize] at (1.5, -0.5) {$d-1$,\\ 12:00};
		\node [align = center, above, font = \scriptsize] at (1.5,1) {Day-Ahead\\Auction};
		\draw [line width = 1] (3,1) -- (3, -0.5);
		\node [align = center, below, font = \scriptsize] at (3, -0.5) {$d-1$,\\ 15:00};
		\node [align = center, above, font = \scriptsize] at (3,1) {Intraday\\Auction};
		\node [align = center, below right, font = \scriptsize] at (3,0) {Hourly Intraday Continuous};
		\draw [line width = 1] (4.05,1) -- (4.05, -0.5);
		\node [align = center, below, font = \scriptsize] at (4.05, -0.5) {$d-1$,\\ 16:00};
		\node [align = center, above right, font = \scriptsize] at (4.05,0) {Quarter-Hourly\\ Intraday Continuous};
		
		\draw [line width = 1] (7,1) -- (7, -0.5);
		\node [align = center, below, font = \scriptsize] at (7, -0.5) {$d-1$,\\ 22:00};
		\node [align = center, above, font = \scriptsize] at  (7,1) {XBID\\ starts};
		
		\draw [line width = 1] (10,1) -- (10, -0.5);
		\node [align = center, below, font = \scriptsize] at (10, -0.5)  {$d$,\\ $s - 60$ min};
		\node [align = center, above, font = \scriptsize] at  (10,1) {XBID\\ closes};
		
		\draw [line width = 1] (11.5,1) -- (11.5, -0.5);
		\node [align = center, above, font = \scriptsize] at (11.5,1) {Market\\ closes};
		\node [align = center, below, font = \scriptsize] at (11.5, -0.5) {$d$,\\ $s - 30$ min};
		
		\draw [line width = 1] (13,1) -- (13, -0.5);
		\node [align = center, above, font = \scriptsize] at (13,1) {Control zones \\ close};
		\node [align = center, below, font = \scriptsize] at (13, -0.5) {$d$,\\ $s - 5$ min};
		\draw [line width = 1] (14,1) -- (14, -0.5);
		\node [align = center, above, font = \scriptsize] at (14,1) {Delivery};
		\node [align = center, below, font = \scriptsize] at (14, -0.5) {$d$, $s$};
		\end{tikzpicture}
		\caption{The daily routine of the German spot electricity market. $d$ corresponds to the day of the delivery and $s$ corresponds to the hour of the delivery.}
		\label{fig:market}
	\end{figure*}

	The German Intraday Continuous Market allows to trade hourly, half-hourly and quarter-hourly products. We conduct the study using the most liquid part of the market -- the hourly one. This is in line with other EPF studies in intraday markets. Trading of hourly products in the German Intraday Continuous begins every day at 15:00 for the 24 products of the following day. It is possible to trade the electricity until 30 minutes (in the whole market) and up to 5 minutes (within respective control zones) before the delivery. In the meantime, between hour 22:00 and 60 minutes before the delivery the cross-border trading within XBID system is possible \citet{kath2019modeling}. This system went live on 18th June 2018. A visualization of the trading timeline %description 
	can be seen in Figure \ref{fig:market}. For more details on the German electricity market, we recommend the paper of \citet{Viehmann2017}.
	
	The most important price measure in the German intraday market is the volume-weighted average price of transactions in the last three hours of trading, called ID$_3$. The index takes into account only these transactions that happen until the gate closure 30 minutes before the delivery, so in fact it measures the last two and a half hours of trading before the gate closure. The relevance of ID$_3$ is an outcome of the behaviour of traders in the intraday market -- most of the transactions are held in this time period making it very liquid. This results in a high interest of practitioners and researchers in the ID$_3$-Price. For more details on the index visit the webpage of EPEX SPOT or see e.g. \citet{narajewski2019econometric}. 
	
	To measure the prices during the trading period, we use the $_x$ID$_y$ defined by \citet{narajewski2019econometric}.
% 	\footnote{
% 	We apply a minor adjustment of the definition % in calculation 
% 	in the case of no trades in the measured time interval. 
% }
% 	with a 
% 	slight difference in calculation 
% 	in the case of no trades in the measured time interval. 
	Let us recall the definition of $_x$ID$_y$. Let $b(d,s)$ be the start of the delivery of a product $s$ on day $d$. By $\mathbb{T}_{x,y}^{d,s} = \left[b(d,s) - x - y, b(d,s) - x \right)$, $x \ge 0$ and $y > 0$, we denote the time interval between $x+y$ and $x$ minutes before the delivery, and by $\mathcal{T}^{d,s}$ we denote a set of timestamps of transactions on the product. The $_x$ID$_y$ is defined by
	
	\begin{equation}
	{}^{}_x\text{ID}_y^{d,s} := \frac{1}{\sum_{k \in \mathbb{T}_{x,y}^{d,s}\cap \mathcal{T}^{d,s}} V_k^{d,s}} \sum_{k \in \mathbb{T}_{x,y}^{d,s}\cap \mathcal{T}^{d,s}} V_k^{d,s}P_k^{d,s},	
	\label{eq:xIDy}
	\end{equation}
	where $V_k^{d,s}$ and $P_k^{d,s}$ are the volume and the price of $k$-th trade within the transaction set $\mathbb{T}_{x,y}^{d,s}\cap \mathcal{T}^{d,s}$ respectively. Let us note that the $_x$ID$_y$ is simply a volume-weighted average price of transactions in the time interval of length $y$ hours and ending $x$ hours before the delivery.
	
	In the case of $\mathbb{T}_{x,y}^{d,s}\cap \mathcal{T}^{d,s} = \emptyset$ we use the value of $_{x+y}$ID$_y$, that is to say the previous observed volume-weighted average price measured on the time period of the same length.\footnote{In \citet{narajewski2019econometric} this value is set to the price of the last transaction. This adjustment is caused by the fact that in this paper we work with 5-minute time intervals, leading to a significant number of the events of no trade in the time interval. This would often result in an artificial change of the price, compared to the previously observed $_x$ID$_y$.}
	In the case of no trades appearing since the start of trading, the price is set to the price of the corresponding Day-Ahead Auction. 
	
	\section{Data and descriptive statistics}
	
	The data used in purpose of this study consists of all transactions on hourly products in the German Intraday Continuous Market between 16th July 2015 and 1st October 2019. A more general descriptive statistics were presented by \citet{narajewski2019econometric}. As mentioned in the previous section, the XBID system started to function on 18th June 2018. This means that XBID trades were possible only on around 30\% of the days in the data. 
	In the forecasting study, we use $D = 365$ days of the data as in-sample, and therefore the analysis in this section is based only on the initial in-sample, i.e. the data between 16th July 2015 and 14th July 2016. The start of the data is set to the first day of lead change in Germany from 45 min to 30 min in order to avoid this structural break.
	In this paper, we aggregate the transactions using the $_x\text{ID}_y^{d,s}$ with $y = 5$ min, and this way we obtain dense time series data. As said before, we are particularly interested in the evolution of prices during the last 2.5 hours of trading before the gate closure, so we use $x \in \mathcal{J} = \{180, 175, \dots, 35, 30 \}$, where $x$ is denoted in minutes. This way we observe $T = 31$ price points a day, what results in $T D = 31 \times 365 = 11315$ in-sample observations and $T$-dimensional simulated trajectories. %In the following paper
	Subsequently, we use a very specific setting, but it can be applied to any other continuous intraday market with other input variables.
	
\begin{figure}[b!]
	 	
	 	\includegraphics[width=1\textwidth]{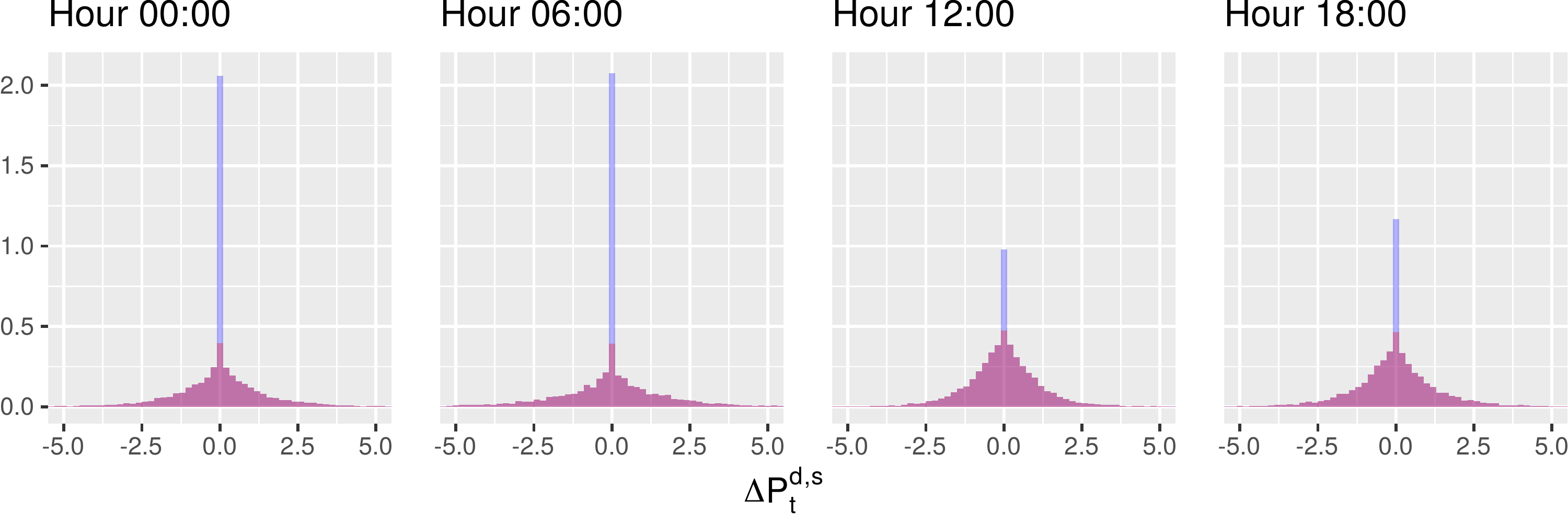}
	 	
	 	\caption{Histograms of the initial in-sample price differences $\Delta P_t^{d,s}$ for selected hours. Blue colour corresponds to the no-trade cases.}
	 	\label{fig:histograms}
	 	
	 \end{figure}

\begin{figure}[t!]
	
	\includegraphics[width=1\textwidth]{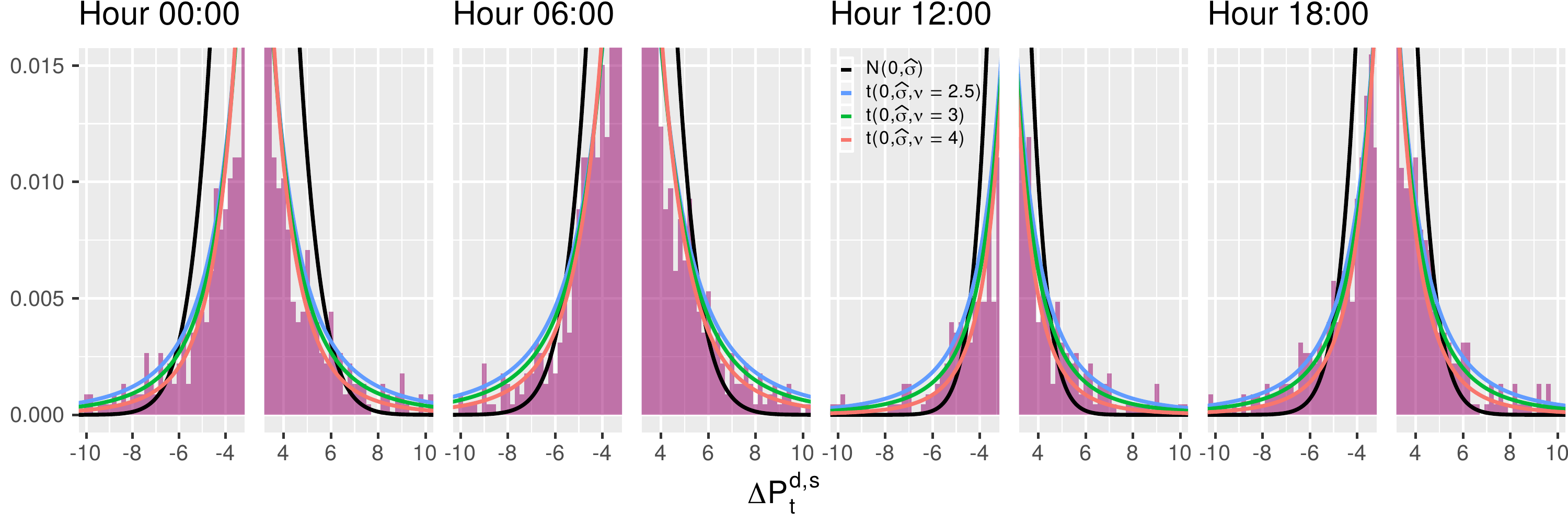}
	\caption{Histograms of the tails of the initial in-sample price differences $\Delta P_t^{d,s}$ for selected hours. Solid lines depict densities of the distributions according to the legend.}
	\label{fig:tails}
	
\end{figure}

	As the market shows strong indications of weak-form efficiency,
	we focus on modelling of the price differences $\Delta P_t^{d,s} = {}_{(T-t) y + 30}\text{ID}_y^{d,s} -{}_{(T-(t-1)) y + 30}\text{ID}_y^{d,s} $ instead of pure prices $P_t^{d,s} = {}_{(T-t) y + 30}\text{ID}_y^{d,s}$. We also introduce the $P_t^{d,s}$ notation for simplicity. Due to the usage of price differences and to the fact that the data is aggregated using 5-minutes grid, we observe a high frequency of observations with no trade, and thus price differences equal to 0. This is depicted in Figure~\ref{fig:histograms}. One can see that lack of transactions happens more often to the night and morning hours. In Figure~\ref{fig:tails}, we zoom in the tails of the histograms from Figure~\ref{fig:histograms}. We also plot there densities of 4 distributions fitted to the data: the normal distribution $\mathcal{N}(0, \widehat{\sigma})$ and the t-distribution $\text{t}(0, \widehat{\sigma}, \nu)$ with fixed $\nu \in \{2.5,3,4\}$ and estimated $\widehat{\sigma}$ using maximum likelihood estimation ignoring the no-trade observations. 
	%The no-trade observations were not utilized in the estimation.
	Based on Figure~\ref{fig:tails} it is clear that the price differences $\Delta P_t^{d,s}$ are heavy-tailed. One can see that even the t-distribution with $\nu = 4$ seems to be not heavy-tailed enough for the data. This indicates that the tail-index of the price differences may be lower than 4 which would mean that the fourth moment of the $\Delta P_t^{d,s}$ might not exist what is a strong indication for heavy tails.
	 
\begin{figure}[t!]
	
	\includegraphics[width=1\textwidth]{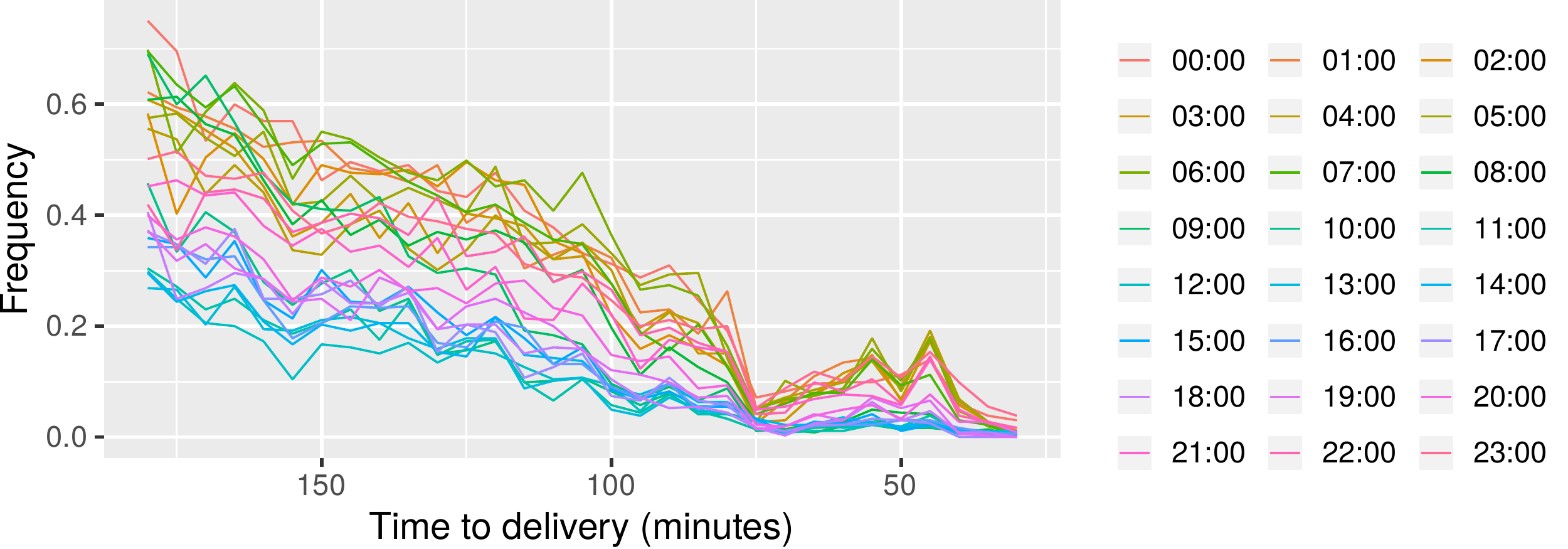}
	
	\caption{Frequency of no-trade event in the initial in-sample price differences $\Delta P_t^{d,s}$ over time to delivery for all 24 products.}
	\label{fig:frequency}
	\end{figure}

	Figure~\ref{fig:frequency} shows the frequency of the no-trade event over time to delivery. We see that the overall behaviour is very similar across all products -- the closer to the delivery, the less observations without transactions. What is different among the products is the level of the frequency. It is clear that the frequency decreases as the product time increases and the reason for it may be the time distance from the Day-Ahead and Intraday Auctions. It is intuitive that since these auctions the uncertainty could be smaller for the first products and higher for the last ones, but the smallest values of frequency are achieved not for the evening, but for the day-peak hours. This can be explained by higher activity in the market due to higher expected demand.%following higher expected load.

	\begin{figure}[t!]
	
	\includegraphics[width=1\textwidth]{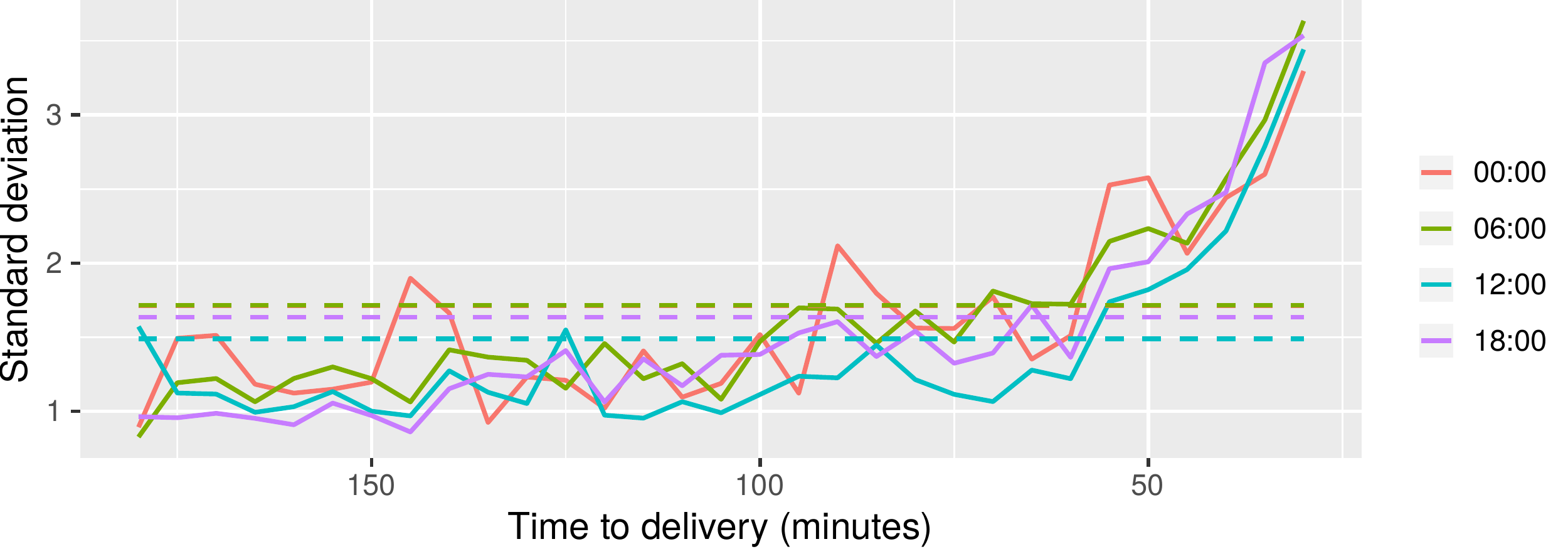}
	
	\caption{Standard deviation of the initial in-sample price differences $\Delta P_t^{d,s}$ for selected hours over time to delivery. Dashed lines indicate the standard deviation independent of time.}
	\label{fig:stdev}
	\end{figure}

	\begin{figure}[t!]
		
		\includegraphics[width=1\textwidth]{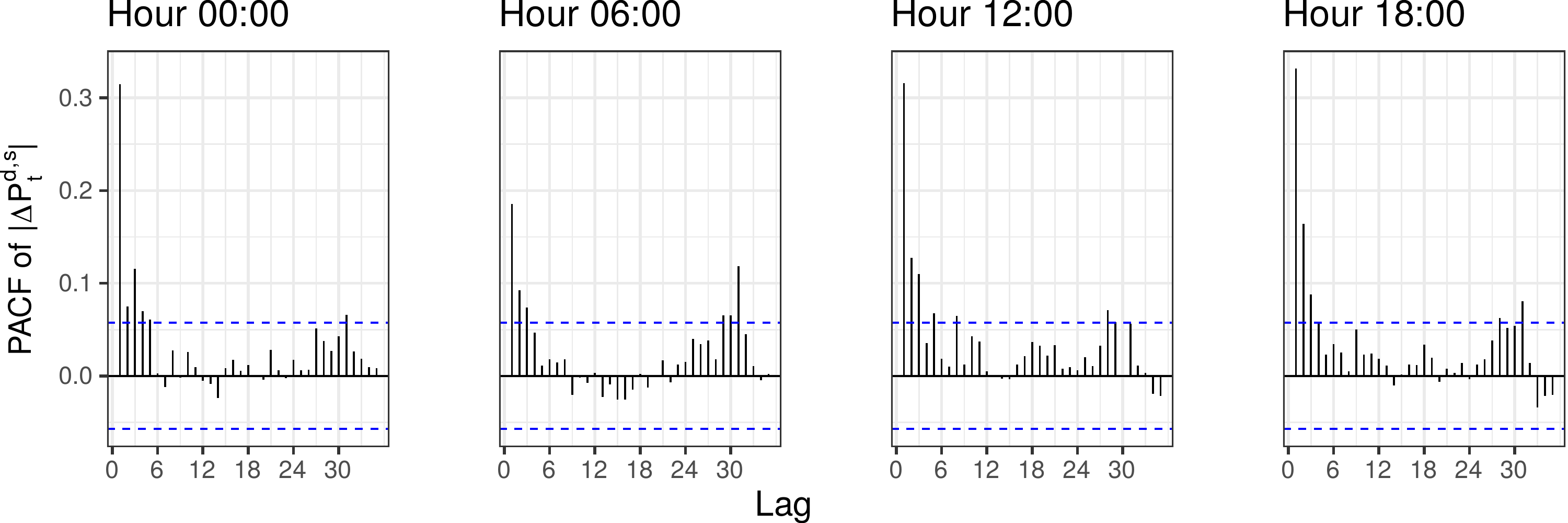}
		
		\caption{Partial autocorrelation function of the initial in-sample absolute price differences $\left|\Delta P_t^{d,s}\right|$ for selected hours. Blue, dashed lines indicate the confidence intervals.}
		\label{fig:pacf}		
	\end{figure}
	Figure \ref{fig:stdev} shows the in-sample standard deviation of price differences $\Delta P_t^{d,s}$ over time to delivery. The dashed lines depict the standard deviation of the whole samples, independent of time. If the price processes would be similar to random walk, the sample standard deviation over time should be oscillating around these dashed lines. The behaviour in Figure \ref{fig:stdev} is clearly different, with a spike in the last 30 minutes before gate closure. This suggests that the variance should be a subject of modelling. Figure~\ref{fig:pacf} presents the partial autocorrelation function of the absolute price differences $\left|\Delta P_t^{d,s}\right|$ to explore potential conditional heteroscedasticity in the heavy-tailed data.
% 	In a standard approach we would inspect the squares instead of absolute values to decide which lags may be informative for the modelling of the volatility. 
% 	However, as we deal with heavy-tailed data, what is depicted in Figure~\ref{fig:tails}, we use the absolute values in purpose of the robustness. 
Figure~\ref{fig:pacf} shows that the most significant are the first three lags. Also, lags up to 6 may contain some information. Surprisingly, lags around 31 seem to be significant too, but this is most likely some daily dependence.
	
	\section{Modelling and estimation}\label{sec:models}

	We assume the price differences $\Delta P_t^{d,s}$ to follow a  4-parametric distribution -- a mixture of the Dirac $\delta_0$ distribution and the 3-parametric t-distribution, sometimes referred as zero-inflated t-distribution:
	\begin{equation}
	G_t^{d,s} = (1-\alpha_t^{d,s})\delta_0 + \alpha_t^{d,s} F_t^{d,s}
	\label{eq:mixture_model}
	\end{equation}
	where $\alpha_t^{d,s} = \mathds{1}(V_t^{d,s} \neq 0)$ is a Bernoulli variable of the event that there is non-zero volume of energy traded on product $s$ on day $d$ at time $t$ with probability $\pi_t^{d,s}$ and $F_t^{d,s}$ is the 3-parametric t-distribution $t(\mu_t^{d,s}, \sigma_t^{d,s}, \nu_t^{d,s})$ where $\mu_t^{d,s} \in \mathbb{R}$ is the mean, $\sigma_t^{d,s} > 0$ the standard deviation and $\nu_t^{d,s} > 2$ the degrees of freedom. The t-distribution is estimated with GAMLSS framework \cite{rigby2005generalized}. %Let us note that if $X \sim F_t^{d,s}$, then $\mathbb{E}(X) = \mu_t^{d,s}$ and $\text{Var}(X) = (\sigma_t^{d,s})^2$. 
	%On the other hand, if $Y \sim G_t^{d,s}$, then $\mathbb{E}(Y) = \pi_t^{d,s}\mu_t^{d,s}$ and $\text{Var}(Y) = (\pi_t^{d,s} \sigma_t^{d,s})^2$.
	
	The GAMLSS is an expansion of the GAM \cite{hastie1990generalized} and it allows to model not only the expected value of a response variable, but also potentially the higher moments, represented by scale and shape parameters. Namely, let $Y$ be a random variable with a density function $f(y|\Theta)$, where $\Theta$ is a set of up to four distribution parameters. Then each $\theta_i \in \Theta$ may be modelled by
	\begin{equation}
	g_i(\theta_i) = \sum_{j = 1}^{J_i} h_{ji} (x_{ji})
	\end{equation}
	where $g_i$ is some link function, $J_i$ is a number of explanatory variables and $h_{ji}$ is a smooth function of explanatory variable $x_{ji}$. Note that function $h_{ji}$ does not have to be a parametric function. In our exercise, we use the following link functions
	\begin{equation}
	\begin{aligned}
	g_1(\mu) = & \mu\\
	g_2(\sigma) = & \log(\sigma) \mathds{1}(\sigma \leq 1) + (\sigma-1) \mathds{1}(\sigma > 1)\\
	g_3(\nu) = & \log(\nu-2).
	\end{aligned}
	\end{equation}	
	\begin{wrapfigure}[10]{r}{0.42\textwidth} 
		\vspace{-0.5cm}
		\includegraphics[width = 1\linewidth]{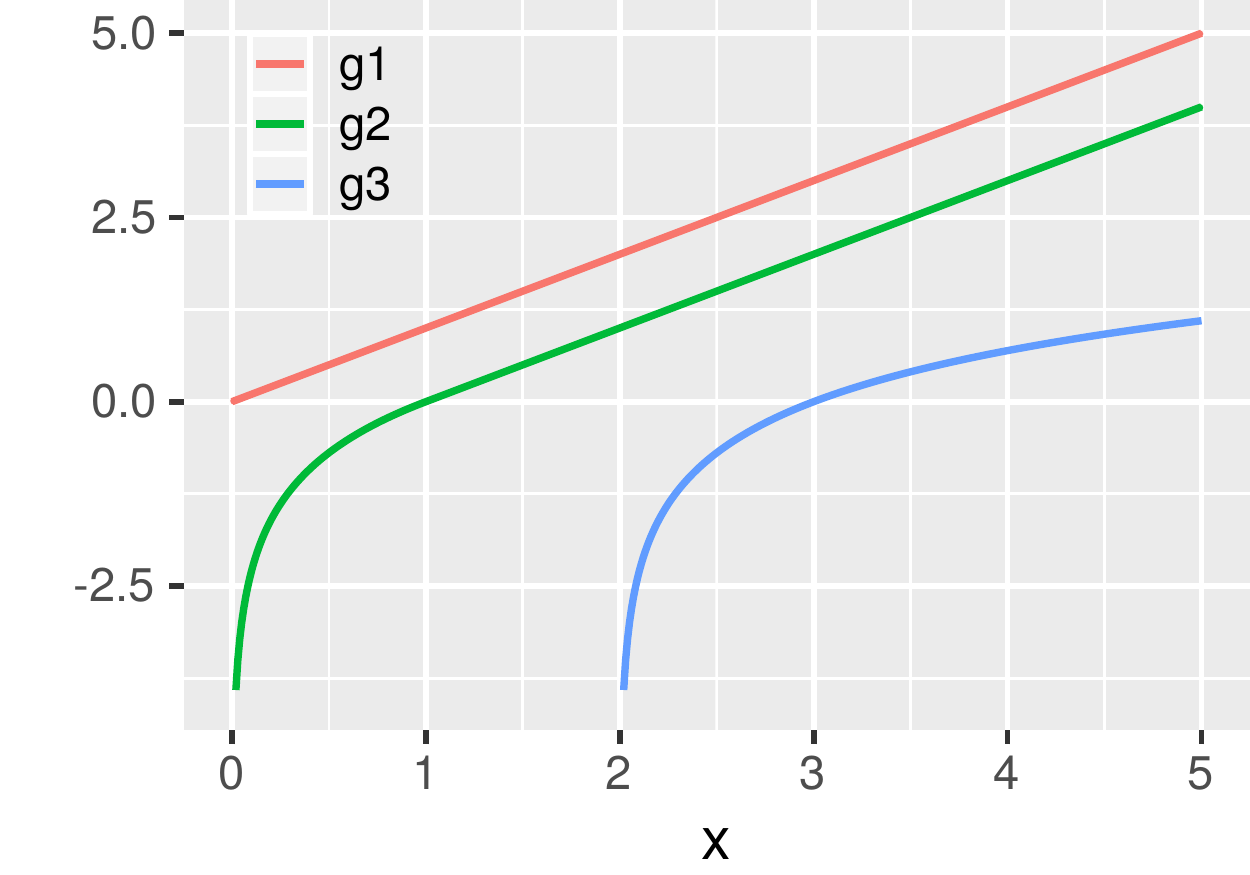}
		\caption{An illustration of the three link functions $g_1$, $g_2$ and $g_3$.}
		\label{fig:link_functions}
	\end{wrapfigure}
	$g_1$ is a standard link function for the expected value. $g_2$ is a link function that we call "logident" and we introduce it in order to avoid exponential inverse function for high values of estimates. The third link function is simply a natural logarithm shifted to 2 for preserving the condition that $\nu > 2$. The three link functions are plotted in Figure~\ref{fig:link_functions}.
	The models for $F_t^{d,s}$ are estimated using the \verb+gamlss+ package in R \cite{stasinopoulos2007generalized}.

	Due to the novelty of the exercise, we cannot use any literature benchmarks, as well as any standard approaches to the modelling of volatility, e.g. GARCH. Even though the data looks like time series, the biggest problem lies in the gap between days. We model each product separately, and for each product we have 31 observations every day. In the corresponding time series, the observations on day $d$ appear in 5-minute breaks, while the time difference between the last observation on day $d$ and the first on day $d+1$ is around 21 hours.

Furthermore, there is no direct link between the prices on day $d$ and day $d+1$ as they are for different delivery periods with potentially different fundamental market situations.	
Thus, the usage of GARCH-type components to address conditional heteroscedasticity is not straight-forward. Instead, as simple benchmarks we use models that assume the distribution of $\Delta \mathbf{P}^{d,s} = (\Delta P_1^{d,s}, \Delta P_2^{d,s}, \dots, \Delta P_T^{d,s})$ to be multivariate, random walk models,  and a model that uses in-sample price differences to create an ensemble forecast. Also, as advanced benchmarks linear quantile regression with copula models are considered.

In the following subsection, the more complicated models are considered. We model explicitly the probability of non-zero number of transactions, the mean, and the variance of fitted distribution. We present the models from the least to the most complex and show the results similarly. This allows us to observe the gain caused by every new part of the model. 
	\subsection{Mixture models}

	We introduce a dependency structure between the first three parameters of the $G_t^{d,s}$ distribution, i.e. $\pi_t^{d,s}$, $\mu_t^{d,s}$ and $\sigma_t^{d,s}$, and the data. For the fourth parameter, the degrees of freedom $\nu_t^{d,s}$, we assume the constancy.
	The $G_t^{d,s}$ distribution is estimated in a 2-step approach. First, the $\pi_t^{d,s}$ parameter is estimated, and then the $F_t^{d,s}$ distribution is fitted to the in-sample price differences $\Delta P_t^{d,s}$ for which the value of $\alpha_t^{d,s}$ is 1.
	
	In the first step, we build a logistic model for $\pi_t^{d,s}$
	\begin{equation}
	\begin{aligned}
	\log\left(\frac{\pi_t^{d,s}}{1-\pi_t^{d,s}}\right) = & \beta_0 + \underbrace{\sum_{j=1}^{3} \beta_{j} \Delta P_{t-j}^{d,s}   + \sum_{j=1}^{6} \beta_{3+j} \left|\Delta P_{t-j}^{d,s} \right| 
	+ \beta_{10} \sum_{j=7}^{12}  \left|\Delta P_{t-j}^{d,s} \right|  }_{\text{price differences}} \\ + & \underbrace{\beta_{11}\text{Mon}(d) +\beta_{12}\text{Sat}(d) + \beta_{13}\text{Sun}(d) 
	+ \sum_{j = 1}^{31} \beta_{13+j} \text{TtM}_j(t)}_{\text{time dummies}} \\ + & \underbrace{\beta_{45}   \text{DA}_{\text{Load}}^{d,s} + \beta_{46}   \text{DA}_{\text{Sol}}^{d,s}   + \beta_{47}   \text{DA}_{\text{WiOn}}^{d,s} + \beta_{48}   \text{DA}_{\text{WiOff}}^{d,s}}_{\text{fundamental regressors}}  + \underbrace{\sum_{j = 1}^{12} \beta_{48+j} \bar{\alpha}_{t-j}^{d,s}}_{\text{regression on $\alpha_t^{d,s}$}}.
	\end{aligned}
	\label{eq:logit_model}
	\end{equation}
	The model explains the logit function with 4 main components: price difference impact, time dummies, fundamental regressors and regression on $\alpha_t^{d,s}$. Price difference impact consists of 3 most recent price differences, 6 most recent absolute price differences and a sum of absolute prices differences lagged by 7 to 12. This component addresses the overall impact of price volatility on $\pi_t^{d,s}$. We expect to observe more trades when the prices are more volatile. Time dummies consist of three weekday dummies and time to maturity dummies. The weekday dummies for Monday, Saturday and Sunday are chosen literature-based. A number of studies \cite{misiorek2006point, uniejewski2016automated, ziel2018day} have proven that usage of these dummies in EPF substantially improves the forecasting performance. These three dummies indicate the end of the week with Monday being a transition day. The use of time to maturity dummies is clear when we take a look again at Figure \ref{fig:frequency}. It is expected that $\pi_t^{d,s}$ rises as we approach the gate closure. Fundamental regressors consist of day-ahead forecasts of total load, solar generation, wind onshore generation and wind offshore generation. It is expected that higher load and share of renewables should rise the uncertainty in the market, and encourage market participants to trade more. The last, but not the least is the regression on $\alpha_t^{d,s}$. We do not use the regression directly, but instead we use the average of last $j$ observed values of $\alpha_t^{d,s}$ which we denote by 
	$\bar{\alpha}_{t-j}^{d,s}$. We expect these values to have a significant impact on the prediction of $\pi_t^{d,s}$. Intuitively, the higher these averages, the higher the value of $\pi_t^{d,s}$.
	
	Model \eqref{eq:logit_model} consists of 61 regressors in total. To avoid overfitting problems, we estimate the model using the least absolute shrinkage and selection operator (lasso) of \citet{Tibshirani1996}. Let us recall that if we possess a logistic model $	\log\left(\frac{\pi}{1-\pi}\right) = \mathbf{X}' \boldsymbol{\beta}$ for the Bernoulli variable $\alpha$ with $P(\alpha = 1) = \pi$, then the lasso estimator $\widehat{\boldsymbol{\beta}}^{\text{lasso}}$ is given by
	\begin{equation}
	\widehat{\boldsymbol{\beta}}^{\text{lasso}} = \arg\min_{\boldsymbol{\beta}} \left\{ - l \left(\boldsymbol{\beta}, \widetilde{\mathbf{X}}\right) + \lambda \left( ||\boldsymbol{\beta}||_1  - |\beta_0| \right) \right\},
	\label{eq:lasso_def}
	\end{equation}
	where $l$ is the corresponding log-likelihood
	\begin{equation}
		l(\boldsymbol{\beta}, \widetilde{\mathbf{X}}) = \frac{1}{N} \sum_{i=1}^{N} \alpha_i  \widetilde{\mathbf{X}}_i' \boldsymbol{\beta} - \log(1 + e^{ \widetilde{\mathbf{X}}_i' \boldsymbol{\beta}}),
	\end{equation}
	$\widetilde{\mathbf{X}}$ is a standardization of $\mathbf{X}$ and $\lambda$ is a tunable shrinkage parameter.
	This method found already many successful applications to the EPF and intraday markets \cite{ziel2016forecasting,uniejewski2019understanding, narajewski2019econometric}. In this exercise, we utilize the \verb+glmnet+ package in R by \citet{friedman2010regularization}. The estimation is conducted using a BIC-tuned $\lambda$ value chosen from an exponential grid of 100 values.
	
\begin{figure}[t!]
		
		\includegraphics[width=1\textwidth]{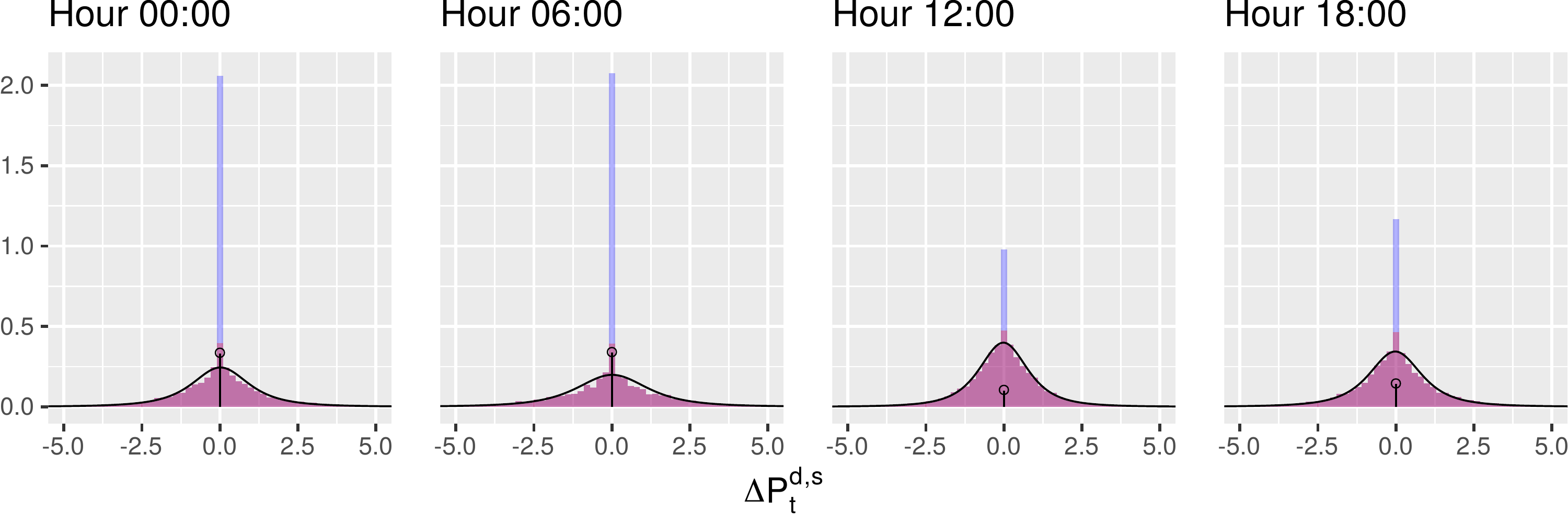}
		
		\caption{Histograms of the initial in-sample price differences $\Delta P_t^{d,s}$ with fitted densities of model \textbf{Mix.RW.t} for selected hours. Blue colour corresponds to the no-trade cases.}
		\label{fig:fitted_densities}
		
	\end{figure}

	Let us now take a look at the $F_t^{d,s}$ distribution in equation \eqref{eq:mixture_model}. We consider four versions of it. In the first one, we assume 
%	$F_t^{d,s} = t(0, \sigma_t^{d,s}, \nu_t^{d,s})$ 
that $F_t^{d,s}$ follows $ t(0, \sigma_t^{d,s}, \nu_t^{d,s})$
	with constant $\sigma_t^{d,s}$ and $\nu_t^{d,s}$. 
	%That is to say, the first mixture model is adjusted \textbf{RW.t.mix.D}. 
We denote it simply by \textbf{Mix.RW.t}. The $F_t^{d,s}$ distribution is fitted to the in-sample price differences with non-zero transaction number using the GAMLSS. With this model we can observe the gain of using a complex model for the $\pi_t^{d,s}$ parameter.
	Figure~\ref{fig:fitted_densities} shows fitted densities to the histograms presented in Figure~\ref{fig:histograms}. They were obtained with model \textbf{Mix.RW.t}.

	The second model utilizes $F_t^{d,s}$ with modelled $\mu_t^{d,s}$ and constant $\sigma_t^{d,s}$ and $\nu_t^{d,s}$, and we denote it by \textbf{Mix.t.mu}. This model helps us understand the outcome of modelling of the expected value of $\Delta P_t$. However, a preliminary analysis has shown that most of the regressors used in model~\eqref{eq:logit_model} were not significant for modelling of $\mu_t^{d,s}$. The only significant were the three most recent price differences. Therefore, we model the expected value with
	
	\begin{equation}
	g_1(\mu_t^{d,s}) = \beta_1 \Delta P_{t-1}^{d,s} + \beta_2 \Delta P_{t-2}^{d,s} + \beta_3 \Delta P_{t-3}^{d,s}.
	\label{eq:mu_formula}
	\end{equation}

	The next model uses $F_t^{d,s}$ with $\mu_t^{d,s} \equiv 0 $, modelled $\sigma_t^{d,s}$ and constant $\nu_t^{d,s}$. We denote it by \textbf{Mix.t.sigma}. The formula for the standard deviation is as follows
	\begin{equation}
	\begin{aligned}
	g_2(\sigma_t^{d,s}) = & \beta_0 + \underbrace{\sum_{j=1}^{6} \beta_{j} \left|\Delta P_{t-j}^{d,s} \right| 
	+ \beta_{7} \sum_{j=7}^{12}  \left|\Delta P_{t-j}^{d,s} \right|}_{\text{absolute price differences}}  + \underbrace{\beta_{8}\text{Mon}(d) +\beta_{9}\text{Sat}(d) + \beta_{10}\text{Sun}(d)}_{\text{weekday dummies}}
	\\+& \underbrace{\beta_{11}   \text{DA}_{\text{Load}}^{d,s} + \beta_{12}   \text{DA}_{\text{Sol}}^{d,s}   + \beta_{13}   \text{DA}_{\text{WiOn}}^{d,s} + \beta_{14}   \text{DA}_{\text{WiOff}}^{d,s}}_{\text{fundamental regressors}} + \underbrace{\beta_{15} \alpha_{t-1}^{d,s} + \beta_{16} \alpha_{t-2}^{d,s}}_{\text{lagged $\alpha_t^{d,s}$}} \\+& \underbrace{h_1(P_{t-1}^{d,s}) + h_2(t)}_{\text{non-linear effects}}
	\end{aligned}
	\label{eq:sigma_model}
	\end{equation}
	
	where $h_1$ and $h_2$ are smooth non-linear P-spline functions. The P-splines simply combine equally-spaced B-splines and discrete penalties. More information on P-splines can be found in \citet{eilers2015twenty}. Let us note that the model described by equation \eqref{eq:sigma_model} uses much more regressors than in equation \eqref{eq:mu_formula}. The explanation of the choice of the variables is very similar to the one of the model described by equation \eqref{eq:logit_model}. We explain the standard deviation of price differences with: lagged absolute price differences,  weekday dummies, fundamental regressors, lagged values of $\alpha_t^{d,s}$ and non-linearities in most recent price and time to maturity variables. 
 	We expect that the absolute price changes are a suitable explanatory variable for the standard deviation as motivated through Figure \ref{fig:pacf}.
% \FZC{	Das kommt unten nochmal. 
% We expect that the price level in the most recent prices is a good explanatory variable for the standard deviation. 
% }
	The fundamental regressors are supposed to have a positive linear correlation with the $\sigma_t^{d,s}$.
	For the Saturday and Sunday dummies we might expect a negative impact due to lower trading activity on weekends, but also a positive impact due to the fact that higher bid-ask spreads are plausible.
	The lagged values of $\alpha_t^{d,s}$ indicate if the market participants traded lately, and thus we believe that it could identify higher price difference's variance. The last two regressors are expected to have a non-linear impact on the formation of $\sigma_t^{d,s}$, and therefore they are estimated using P-splines. Figure \ref{fig:stdev} provides already an evidence that the standard deviation varies over time to maturity. Moreover, we suspect that extreme values of most recent price $P_{t-1}^{d,s}$ result in a higher variance due to a relatively inelastic supply curve in extreme price areas.
	
	The last and at the same point the most complicated model uses $F_t^{d,s}$ with $\mu_t^{d,s}$ and $\sigma_t^{d,s}$ modelled and constant $\nu_t^{d,s}$. We denote it by \textbf{Mix.t.mu.sigma}. The $\mu_t^{d,s}$ is modelled using the formula from equation \eqref{eq:mu_formula} and the $\sigma_t^{d,s}$ using the formula from equation \eqref{eq:sigma_model}.	
	Let us mention that we could make the mixture model even more complex by modelling the degrees of freedom parameter $\nu_t^{d,s}$. However, a preliminary analysis has shown that it does not yield any significant improvement while increasing heavily the computational cost. Thus, in the forecasting study we analyse the performance of 8 models described in this section.
	
		\subsection{Simple benchmark models}
	The first benchmark model uses one of $D = 365$ historical trajectories to model the price difference vector $\Delta \mathbf{P}^{d,s} = (\Delta P_1^{d,s}, \Delta P_2^{d,s}, \dots, \Delta P_T^{d,s})$. We denote it by \textbf{Naive} and its formula is given by 
		\begin{equation}
		\Delta \mathbf{P}^{d,s} = \Delta  \mathbf{P}^{d',s}
		\end{equation}
	where $d' \sim \mathcal{U}(\{d-1,\ldots, d-D\})$ is a uniform random variable indicating the day used to model the price difference. Let us note that a fixed $d'$ index is used to model the whole price trajectory, i.e. for every $t \in \{1,2,\dots, T\}.$ This model assumes that the future trajectories can be forecasted using simply the past ones.
	
The second and the third benchmark models assume that the price difference vector  $\Delta \mathbf{P}^{d,s}$ follows a multivariate normal and t-distributions, respectively.
	They are denoted by \textbf{MV.N} and \textbf{MV.t} and are given by
	
	\begin{equation}
	\Delta \mathbf{P}^{d,s} = \boldsymbol{\varepsilon}^{d,s} 
	\end{equation}
 
	where $\boldsymbol{\varepsilon}^{d,s} \sim \mathcal{N}\left(\mathbf{0}, \boldsymbol{\Sigma}^{d,s}\right)$ in the case of $\mathbf{MV.N}$ and $\boldsymbol{\varepsilon}^{d,s} \sim t \left(\mathbf{0}, \boldsymbol{\Sigma}^{d,s}, {\nu}^{d,s} \right)$ in the case of \textbf{MV.t}. Let us note that the covariance matrix ${\boldsymbol{\Sigma}}^{d,s}$ and degrees of freedom $\nu^{d,s}$  are estimated by fitting the respective distributions to the in-sample observations. Moreover, the degrees of freedom $\nu^{d,s}$ is assumed to be constant for all  $t \in \{1,2,\dots, T\}.$

	The next benchmark model is the random walk version of the mixture model described by equation \eqref{eq:mixture_model}, and we denote it by \textbf{RW.t.mix.D}. The formula is as follows
	\begin{equation}
	\Delta P_t^{d,s} = P_t^{d,s} -  P_{t-1}^{d,s} = \varepsilon_t^{d,s}
	\label{eq:RW.t.mix.D}
	\end{equation}
	where $\varepsilon_t^{d,s} \sim G_t^{d,s}$ with $\widehat{\pi}_t^{d,s} = \frac{1}{D T} \sum_{i = d - D}^{d-1} \sum_{t=1}^{T} \mathds{1}(V_t^{i,s} \neq 0)$, ${\mu}_t^{d,s} \equiv 0$ and constant ${\sigma}_t^{d,s}$ and ${\nu}_t^{d,s}$. These values are estimated based on the in-sample data.
	
	The fifth benchmark model is a modification of the \textbf{RW.t.mix.D}. We denote it by \textbf{RW.t}, and we simply set ${\pi}_t^{d,s} \equiv 1$ which means that we do not incorporate the mixing part and assume that the price differences follow the t-distribution. The last and the simplest of the random walk models assumes the price differences to follow a Gaussian distribution $\mathcal{N}(0, ({\sigma}_t^{d,s})^2)$ and is denoted by \textbf{RW.N}. In terms of the $G_t^{d,s}$ distribution, we simply modify the \textbf{RW.t} model by taking ${\nu}_t^{d,s} \to \infty$.
	
	Later, we consider the random walk models from the simplest \textbf{RW.N} to the most complex \textbf{RW.t.mix.D}. This allows us to observe the gain of introducing more complex structure of the distribution. Let us note that model \textbf{RW.N} assumes exponentially decaying tails of the price differences $\Delta P_t^{d,s}$. Comparing it to model \textbf{RW.t} we measure the gain of assuming heavier, polynomially decaying tails.  Based on the number of outlier observations in the German intraday market and on Figure~\ref{fig:tails}, we expect it to perform better than the Gaussian random walk. Then, considering the \textbf{RW.t.mix.D} helps us to understand the gain of the introduction of the mixture.
	
\subsection{Advanced benchmark models}
	
	As mentioned, we are unable to use any literature-based benchmarks as this is the first paper on ensemble forecasting in intraday electricity markets. However, it is possible to implement scenario generating methods that are utilized in other research areas. Thus, as advanced benchmark we utilize a smoothed linear quantile regression model with two copulas: Gaussian and independence, and we denote them by \textbf{LQR.Gauss} and \textbf{LQR.ind}, respectively. A very similar approach was applied recently in the purpose of generating density forecasts of significant sea wave height and peak wave period~\cite{gilbert2020probabilistic}.
	
	First, we build the linear quantile regression (LQR) model using the same set of regressors as for the mixture models. The formula is as follows
	\begin{equation}
	\begin{aligned}
	Q^{\tau}\left(\Delta P_t^{d,s}\right)= & \beta_0 + \underbrace{\sum_{j=1}^{3} \beta_{j} \Delta P_{1-j}^{d,s}   + \sum_{j=1}^{6} \beta_{3+j} \left|\Delta P_{1-j}^{d,s} \right| 
		+ \beta_{10} \sum_{j=7}^{12}  \left|\Delta P_{1-j}^{d,s} \right| + \beta_{11} P_0^{d,s} }_{\text{price components}} \\ + & \underbrace{\beta_{12}\text{Mon}(d) +\beta_{13}\text{Sat}(d) + \beta_{14}\text{Sun}(d)} + \underbrace{\beta_{15} \alpha_{1-1}^{d,s} + \beta_{16} \alpha_{1-2}^{d,s}}_{\text{lagged $\alpha_t^{d,s}$}} \\ + & \underbrace{\beta_{17}   \text{DA}_{\text{Load}}^{d,s} + \beta_{18}   \text{DA}_{\text{Sol}}^{d,s}   + \beta_{19}   \text{DA}_{\text{WiOn}}^{d,s} + \beta_{20}   \text{DA}_{\text{WiOff}}^{d,s}}_{\text{fundamental regressors}}
	\end{aligned}
	\label{eq:lqr_model}
	\end{equation}
	for $\tau \in  \{0.01, 0.02, \dots, 0.99 \}$ and $t = 1, 2 \dots, T$. That is to say, we build separate models for each quantile $\tau$ and each time point $t$. Let us note that due to the design of the model, we can use only the regressor values available at the time of forecasting $t = 0$ (i.e. 3~h 5~min before the delivery). This results in the fact that here we model all $T$ time points using the same data, what is contrary to the mixture models where we can use autoregressive variables due to the recursive character of the models. We estimate the LQR models using the \verb|quantreg| package in R~\cite{quantreg}.
	
	In the next step, a spline interpolation is applied over all fitted $Q^{\tau}\left(\Delta P_t^{d,s}\right)$ for $\tau \in \{0.01, 0.02, \dots, 0.99 \}$ and for every $t = 1, 2 \dots, T$. In order to preserve the monotonicity of the estimated cumulative distribution function (CDF) we compute a monotonic cubic spline using Hyman filtering~\cite{hyman1983accurate}. This way we obtain a smooth and monotonic $T$-dimensional CDF function
	 \begin{equation}
	 \boldsymbol{\Psi}^{d,s} (\mathbf{x}) = \left(\Psi_1^{d,s}(x_1), \Psi_2^{d,s}(x_2), \dots, \Psi_T^{d,s}(x_T) \right)
	 \end{equation} 
	 where $\Psi_t^{d,s}\left( \min_{\{d-1, \dots, d-D  \}}\left( \Delta P_t^{d,s} \right) \right) = 0$ and $\Psi_t^{d,s}\left( \max_{\{d-1, \dots, d-D  \}}\left( \Delta P_t^{d,s} \right) \right) = 1$. To assess the dependency structure of the price differences $\Delta P_t^{d,s}$ over $t$ we use two copulas: Gaussian and independence. The Gaussian copula for a given correlation matrix $\mathbf{R}$ can be written as
	 \begin{equation}
	 	C_{\mathbf{R}}(\mathbf{u}) = \boldsymbol{\Phi}_\mathbf{R}\left( \Phi^{-1}(u_1), \Phi^{-1}(u_2), \dots, \Phi^{-1} (u_T) \right)
	 \end{equation}
	 where $\Phi^{-1}$ is the inverse CDF of a standard normal distribution and $\boldsymbol{\Phi}_\mathbf{R}$ is a joint CDF of a multivariate normal distribution with mean vector zero and covariance matrix equal to the correlation matrix $\mathbf{R}$. We estimate the correlation matrix $\mathbf{R}$ simply by calculating the in-sample correlation matrix.

	\section{Forecasting study and evaluation}
	
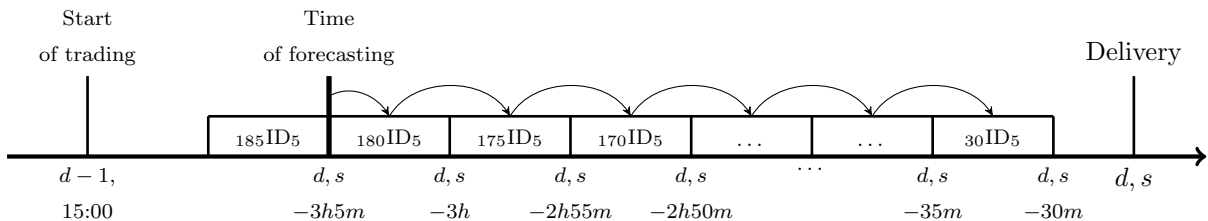
\begin{figure}[b!]
		\centering
		\begin{tikzpicture}[scale=1.07]
		\draw [->] [ultra thick] (0,0) -- (14.9,0);
		
		\node [align = center, below, font = \scriptsize] at (1, 0) {$d-1$,\\ 15:00};
		\node [align = center, below, font = \scriptsize] at (4, 0) {$d, s$\\$-3h5m$};
		\node [align = center, below, font = \scriptsize] at (5.5, 0) {$d,s$\\$ - 3h$};
		\node [align = center, below, font = \scriptsize] at (7, 0) {$d, s$\\$ - 2h55m$};
		\node [align = center, below, font = \scriptsize] at (8.5, 0) {$d,s$\\$ - 2h50m$};
		\node [align = center, below, font = \scriptsize] at (10, 0) {\dots};
		\node [align = center, below, font = \scriptsize] at (11.5, 0) {$d,s$\\$ - 35m$};
		\node [align = center, below, font = \scriptsize] at (13, 0) {$d,s$\\$ - 30m$};
		\node [align = center, below, font = \footnotesize] at (14, 0) {$d,s$};
		
		\draw [line width = 1] (1, 1) -- (1,0);
		
		\draw [line width = 2] (4, 1) -- (4,0);
		\node [align = center, above, font = \scriptsize] at (4,1) {Time \\of forecasting};
		
		\draw [line width = 1] (2.5, 0.5) -- (2.5,0);
		\draw [line width = 1] (5.5,0.5) -- (5.5,0);
		\draw [line width = 1] (7, 0.5) -- (7,0);
		\draw [line width = 1] (8.5,0.5) -- (8.5, 0);
		\draw [line width = 1] (10,0.5) -- (10, 0);
		\draw [line width = 1] (11.5,0.5) -- (11.5, 0);
		\draw [line width = 1] (13,0.5) -- (13, 0);
		\draw [line width = 1] (14,1) -- (14, 0);
		
		\draw [line width = 1] (2.5, 0.5) -- (13,0.5);
		
		\node [align = center, above, font = \scriptsize] at (1,1) {Start \\of trading};
		\node [align = center, above, font = \footnotesize] at (14,1) {Delivery};
		
		\node [align = center, above, font = \scriptsize] at (3.25,-0.01) {$_{185}\text{ID}_{5}$};
		\node [align = center, above, font = \scriptsize] at (4.75,-0.01) {$_{180}\text{ID}_{5}$};
		\node [align = center, above, font = \scriptsize] at (6.25,-0.01) {$_{175}\text{ID}_{5}$};
		\node [align = center, above, font = \scriptsize] at (7.75,-0.01) {$_{170}\text{ID}_{5}$};
		\node [align = center, above, font = \scriptsize] at (9.25,-0.01) {\dots};
		\node [align = center, above, font = \scriptsize] at (10.75,-0.01) {\dots};
		\node [align = center, above, font = \scriptsize] at (12.25,-0.01) {$_{30}\text{ID}_{5}$};
		\draw [->, >=stealth'] (4,0.75)  to [out=30,in=120] (4.75,0.5);
		\draw [->, >=stealth'] (4.75,0.5)  to [out=60,in=120] (6.25,0.5);
		\draw [->, >=stealth'] (6.25,0.5)  to [out=60,in=120] (7.75,0.5);
		\draw [->, >=stealth'] (7.75,0.5)  to [out=60,in=120] (9.25,0.5);
		\draw [->, >=stealth'] (9.25,0.5)  to [out=60,in=120] (10.75,0.5);
		\draw [->, >=stealth'] (10.75,0.5)  to [out=60,in=120] (12.25,0.5);
		\end{tikzpicture}
		\caption{An outline of the ensemble forecasting exercise.}
		\label{fig:forecasting_outline}
	\end{figure}

	We use a rolling window forecasting study approach with $D = 365$ days in-sample size and $N = 1173$ days out-of-sample size. The in-sample data consists of $D  T$ data points where $T=31$ in this study. We model each of the $S = 24$ hourly products separately and our forecasting time is 185 minutes before the delivery of product $s$ on day $D+1$. That is to say, we can utilize all the information from the in-sample data and from the day $D+1$ until 185 minutes before the delivery. At this time we forecast $M = 1000$ times the first price difference $\Delta P_1^{d,s} = {}_{180}\text{ID}_5^{d,s} - {}_{185}\text{ID}_5^{d,s}$. Based on these forecasts and explanatory data we simulate $M$ second price differences $\Delta P_2^{d,s}$ and we continue this recursive process until we reach the gate closure. Figure \ref{fig:forecasting_outline} provides an outline of the exercise. This gives us $M$ simulated trajectories, each consisting of $T=31$ points. After that, we move the window forward by one day and repeat the exercise until the end of out-of-sample data. However, in the case of benchmark models we do not use the recursive algorithm as there is no recursion in their formulas.

Before we discuss the evaluation design in detail, we recall that we are mainly interested in the 
	evaluation of the forecasted $T$-dimensional distribution of the price vector $\mathbf{P}^{d,s} = \left(P_{1}^{d,s},\ldots,P_{T}^{d,s}\right)$ which is represented by the predicted ensemble. Indeed, the multivariate cumulative distribution function of the ensemble 
	coincides with the underlying cumulative distribution function if the ensemble sample size $M$ goes to infinity. Thus, for sufficiently large $M$ the evaluation of the scenario set can be regarded as the evaluation of probabilistic distributions.
	
	From the theoretical point of view, strictly proper multivariate scoring rules are the first choice for evaluation, as they are able to identify the optimal forecast resp. the true distribution, see \citet{gneiting2007strictly, pinson2012evaluating}. However, we want to remind us that even if forecast $A$ performs significantly better than forecast $B$ with respect to a strictly proper scoring rule, there is no guarantee that $A$ also performs better than $B$ in stochastic optimization problems (e.g. trading or storage optimization) where the forecasts are used as input. The optimal forecast would always yield optimal solutions in the stochastic optimization application. Thus, if $A$ is close to the optimal forecast with respect to a strictly proper multivariate scoring rule, the aforementioned risk that $B$ outperforms $A$ in the application is very limited if the stochastic optimization problem is continuous in the stochastic argument. Unfortunately, this only holds for strictly proper multivariate scoring rules. For proper scoring rules which identify only some characteristics of the full predictive distribution, this is certainly not true. 
	The range of available strictly proper scoring rules is very limited, and reduces basically to the energy score for our ensemble forecasting problem, see e.g. \citet{lerch2020simulation}.
	Therefore, we consider also proper scoring rules which might allow further insights as they focus on specific characteristics of the full distribution. 
	To draw statistically significant conclusions on the outperformance of the forecasts of the considered models we utilize also the \citet{diebold1995comparing} test.
	
	As mentioned, the only available multivariate strictly proper scoring rule is the energy score (ES) \cite{gneiting2007strictly}\footnote{Also the multivariate log-score is a known strictly proper scoring rule for multivariate distributions. However, it requires that the underlying multivariate distribution is continuous and has a density. Due to the non-trade events this is not satisfied for our forecasting problem.}.
	We compute the ES loss function in the following way
	\begin{equation}
		\text{ES}^{d,s} = \text{ED}^{d,s} - \frac{1}{2} \text{EI}^{d,s} 
	\end{equation}
	where 
	\begin{equation}
			\text{ED}^{d,s} =  \frac{1}{M} \sum_{j = 1}^{M} \left|\left| \mathbf{P}^{d,s} - \widehat{\mathbf{P}}_j^{d,s}  \right|\right|_2
	\end{equation}
	and
	\begin{equation}
		\text{EI}^{d,s} =  \frac{1}{\frac{1}{2} M(M - 1)} \sum_{j = 1}^{M} \sum_{i = j+1}^{M} \left|\left| \widehat{\mathbf{P}}_j^{d,s}  - \widehat{\mathbf{P}}_i^{d,s}   \right|\right|_2
	\end{equation}
	with $\mathbf{P}^{d,s} = \left( P_1^{d,s}, P_2^{d,s}, \dots, P_T^{d,s} \right)$ and $\widehat{\mathbf{P}}_j^{d,s} = \left(\widehat{P}_{1,j}^{d,s}, \widehat{P}_{2,j}^{d,s}, \dots, \widehat{P}_{T,j}^{d,s}\right)$. The $\text{ED}^{d,s}$ component measures the distance between the simulated trajectories and the observed prices. On the other hand, $\text{EI}^{d,s}$ measures the spread between the simulations. To calculate the overall energy score we use an average
	\begin{equation}
		\text{ES} = \frac{1}{S N } \sum_{s = 1}^{S} \sum_{d = 1}^{N} \text{ES}^{d,s}.
	\end{equation}
	We mentioned that the ES evaluates the full predictive distribution, which includes the path dependency in the generated scenarios. To illustrate the appropriateness of the energy score to evaluate correctly an ensemble forecasting study, we perform a short experiment in the results section. We take the best performing model and modify it with 3 different copulas which we refer as maximum dependency, minimum dependency and independence. 
	For the maximum dependency copula we consider the $T$-dimensional co-monotonicity copula
	defined by $ M_{\max}(\boldsymbol{u}) = \min(u_1 , u_2 ,\ldots, u_T)$.
	The minimum dependency copula $M_{\min}$ is constructed using pairwise bivariate
	counter-monotonicity copulas defined by $W(u_1, u_2) = \max(u_1 + u_2 - 1, 0)$. So $M_{\min}$ is the copula that is associated with the $T$-dimensional uniform random variable $\boldsymbol{U} = (U_1,\ldots, U_T)$ that satisfies $(U_{t},U_{t+1}) \sim W$ for all $t=1,\ldots, T-1$.
	The $T$-dimensional independence copula is defined by $M_{\text{ind}}(\boldsymbol{u}) = \Pi_{t=1}^T u_t$.
	% 	comonotonicity copula M d (u) = min{u 1 , u 2 , . . . , u d } associated with a vector
	% U = (U 1 ,U 2 , . . . ,U d ) of r.v.’s uniformly distributed on I and such that U 1 = U 2 =
	% · · · = U d almost surely
	The 3 new models are evaluated using all considered measures and compared to the original one.

	As pointed out by \citet{pinson2012evaluating}, the ES does not evaluate the ability of the trajectories to mimic specific characteristics of the stochastic process. Therefore, we also focus our evaluation on specific characteristics of the underlying multivariate distribution. In this purpose, we consider additionally the subsequent proper scoring rules.
	We utilize the mean absolute error (MAE) and the root mean squared error (RMSE), pinball score (PB) to evaluate the median, mean and selected quantile trajectories, respectively, see e.g. \cite{gneiting2011making}. For evaluation of the marginal density fit of our scenarios, we consider the continuous ranked probability score (CRPS) and additionally the empirical coverage of specific prediction intervals \cite{gneiting2007strictly}. 
	Moreover, we consider the variogram score (VS) and Dawid-Sebastiani score (DSS) which are regularly used to evaluate multivariate distributions \cite{lerch2020simulation}. Note that both measures are only proper scoring rules and correct model identification fails in general, see e.g. \cite{ziel2019multivariate} for empirical examples. Other scoring rules evaluating e.g. marginal distribution characteristics or specific events might also be added if it is relevant for the desired application.

The RMSE is the optimal least squares measure, i.e. it is the strictly proper scoring rule for mean evaluation while MAE is strictly proper for median evaluation. They are widely used both by researchers and practitioners. Their formulas are given by
	\begin{equation}
		 \text{RMSE}  =  \sqrt{\frac{1}{S N  T} \sum_{s = 1}^{S} \sum_{d = 1}^{N} \sum_{t = 1}^{T}\left(P_t^{d,s} - \frac{1}{M} \sum_{j=1}^{M} \widehat{P}_{t,j}^{d,s}\right)^2}
		\label{eq:rmse}
	\end{equation}
	and
	\begin{equation}
		\text{MAE} =  \frac{1}{S N  T} \sum_{s = 1}^{S} \sum_{d = 1}^{N} \sum_{t = 1}^{T} \left|P_t^{d,s} - \text{med}_{j = 1, \dots, M} \left(\widehat{P}_{t,j}^{d,s} \right) \right|
	\end{equation}
	where $\widehat{P}_{t,j}^{d,s}$ is the $j$-th simulation of $P_t^{d,s}$ and $\text{med}_{j = 1, \dots, M} \left(\widehat{P}_{t,j}^{d,s} \right)$ is the median of $M$ simulated $\widehat{P}_{t,j}^{d,s}$ prices.
	
	We approximate the CRPS using the pinball loss
	\begin{equation}
		\text{CRPS}_t^{d,s} = \frac{1}{R} \sum_{\tau \in r} \text{PB}_{t, \tau}^{d,s}
	\end{equation}
	for a dense equidistant grid of probabilities $r$ between 0 and 1 of size $R$, see e.g.~\cite{nowotarski2018recent}. In this study, we consider $r = \{0.01, 0.02,\dots, 0.99\}$ of size $R = 99$. $\text{PB}_{t,\tau}^{d,s}$ is the pinball loss with respect to probability $\tau$. Its formula is given by
	\begin{equation}
		\text{PB}_{t, \tau}^{d,s} = \left(\tau - \mathds{1}_{\left\{ P_t^{d,s} < Q_{j = 1, \dots, M}^{\tau}\left(\widehat{P}_{t,j}^{d,s}\right)\right\}} \right) \left(P_t^{d,s} - Q_{j = 1, \dots, M}^{\tau}\left(\widehat{P}_{t,j}^{d,s}\right) \right)  
	\end{equation}
	where $Q_{j = 1, \dots, M}^{\tau}\left(\widehat{P}_{t,j}^{d,s}\right)$ is the $\tau$-th quantile of $M$ simulated $\widehat{P}_{t,j}^{d,s}$ prices. To calculate the overall CRPS value we use a simple average
	\begin{equation}
		\text{CRPS} = \frac{1}{S N  T} \sum_{s = 1}^{S} \sum_{d = 1}^{N} \sum_{t = 1}^{T} \text{CRPS}_t^{d,s}.
	\end{equation}
	We can also use the pinball loss to compare the models' performance in particular quantiles. In this purpose the following formula is used
	\begin{equation}
		\text{PB}_{\tau} = \frac{1}{S N  T} \sum_{s = 1}^{S} \sum_{d = 1}^{N} \sum_{t = 1}^{T} \text{PB}_{t,\tau}^{d,s}.
	\end{equation}
As mentioned, we use also the empirical coverage of prediction intervals, precisely the $(\tau/2,1-\tau/2)$-prediction interval. 
The $\tau\%$-coverage is calculated using the following formula
	\begin{equation}
		\tau\%\text{-cov} = \frac{1}{S N T} \sum_{s = 1}^{S} \sum_{d = 1}^{N} \sum_{t = 1}^{T} \mathds{1}_{\left\{ Q_{j = 1, \dots, M}^{(1-\tau)/2}(\widehat{P}_{t,j}^{d,s}) < P_t^{d,s} < Q_{j = 1, \dots, M}^{(1+\tau)/2}(\widehat{P}_{t,j}^{d,s}) \right\} }
	\end{equation}
	where $\tau \in \{0.5, 0.9, 0.99\}$.

	The variogram score was introduced by \citet{scheuerer2015variogram} in a probabilistic forecasting exercise for meteorological data. We compute it by
		\begin{equation}
		\text{VS} = \frac{1}{SNT^2} \sum_{s = 1}^S \sum_{d = 1}^{N} \sum_{i = 1}^{T} \sum_{j = 1}^{T} 
		\left( \left|P_{i}^{d,s}-P_{j}^{d,s}\right| -
		  \sum_{k=1}^M \left|\widehat{P}_{i,k}^{d,s}-\widehat{P}_{j,k}^{d,s}\right|  \right)^2
	\end{equation}
The Dawid-Sebastiani score evaluates the first and second moments \cite{gneiting2011making}. It corresponds to the log-score of the multivariate normal distribution. We calculate it by
		\begin{equation}
		\text{DSS} =  \frac{1}{SN} \sum_{s = 1}^{S}	\sum_{d = 1}^{N} \log\left( \text{det}\left( \widehat{\boldsymbol{\Sigma}}^{d,s}\right)\right) + \left(\mathbf{P}^{d,s} -\widehat{\boldsymbol{\mu}}^{d,s}\right)'\left(\widehat{\boldsymbol{\Sigma}}^{d,s}\right)^{-1}\left(\mathbf{P}^{d,s} - \widehat{\boldsymbol{\mu}}^{d,s}\right)
	\end{equation}
where $\widehat{\boldsymbol{\mu}}^{d,s}$ and $\widehat{\boldsymbol{\Sigma}}^{d,s}$ are the sample mean vector and the sample covariance matrix of the predicted price ensemble.
	
	However, the aforementioned measures do not allow us to make conclusions regarding the statistical significance. To do so we utilize the \citet{diebold1995comparing} test which tests forecasts of model $A$ against the ones of model $B$. 
		The DM test is mostly used to evaluate point forecasts, but with correctly defined loss differential series it can be successfully applied in the evaluation of probability forecasts. We derive the series using ES and CRPS what was already applied by e.g.~\citet{muniain2020probabilistic} and~\citet{lerch2020simulation}.
	We utilize the multivariate version of the DM test as \citet{ziel2018day}. The multivariate DM test results in one statistic for each model which is computed based on the $S$-dimensional vector of losses per day. Therefore, denote $L_A^d = (L_A^{d,1}, L_A^{d,2}, \dots, L_A^{d,S})'$ and $L_B^d = (L_B^{d,1}, L_B^{d,2}, \dots, L_B^{d,S})'$ the vectors of out-of-sample losses for day $d$ of models $A$ and $B$, respectively. By $L_Z^{d,s}$ we mean the $\text{ES}^{d,s}$ and $\text{CRPS}^{d,s}$ losses of model Z, formally we choose 
	\begin{equation}
		L_Z^{d,s} = \text{ES}^{d,s} \quad \text{and} \quad L_Z^{d,s} = \text{CRPS}^{d,s} = \frac{1}{T} \sum_{t=1}^{T} \text{CRPS}_t^{d,s}.
	\end{equation}
	The multivariate loss differential series 
	\begin{equation}
		\Delta_{A,B}^d = ||L_A^d||_1 - ||L_B^d||_1
	\end{equation}
	defines the difference of losses in $||\cdot||_1$ norm. For each model pair, we compute the p-value of two one-sided DM tests. The first one is with the null hypothesis $\mathcal{H}_0: \mathbb{E}(\Delta_{A,B}^d) \leq 0$, that is to say the outperformance of the forecasts of model $B$ by the ones of model $A$. The second test is the reverse null hypothesis $\mathcal{H}_0: \mathbb{E}(\Delta_{A,B}^d) \geq 0$. Let us note that these tests are complementary, and we assume that the loss differential series is covariance stationary.
	
	\section{Results}
	We divided this section into two subsections: in the first one, we inspect the in-sample characteristics and in the second one, we present the out-of-sample simulation results.
	
	\subsection{In-sample characteristics}
	We start our study with an analysis of the initial in-sample characteristics. Table \ref{tab:coefs} shows the estimated coefficient values of model \textbf{Mix.t.mu.sigma} based on the initial in-sample data. The table reports the values for every hourly product, and it is split to 3 sub-tables, each regarding different parameter of the t-distribution. The first sub-table presents coefficients of the model described by equation \ref{eq:mu_formula}. Variable $\Delta P_{t-1}^{d,s}$ appears to be statistically significant for most of the hours. However, raising the lag decreases the significance. This behaviour goes in the direction of weak-form efficiency concluded by \citet{narajewski2019econometric}.

	\begin{table}[t]
		\centering
		\setlength{\tabcolsep}{2pt}
		\begin{adjustbox}{max width=1\textwidth}
		\begin{tabular}{rrrrrrrrrrrrrrrrrrrrrrrrr}
  \hline
$g_1(\mu_t^{d,s})$ & 0 & 1 & 2 & 3 & 4 & 5 & 6 & 7 & 8 & 9 & 10 & 11 & 12 & 13 & 14 & 15 & 16 & 17 & 18 & 19 & 20 & 21 & 22 & 23 \\ 
\hline
$\Delta P_{t-1}^{d,s}$ & \cellcolor[rgb]{0,0.5,0} {-0.09} & \cellcolor[rgb]{0,0.5,0} {-0.09} & \cellcolor[rgb]{0,0.5,0} {-0.11} & \cellcolor[rgb]{0,0.5,0} {-0.09} & \cellcolor[rgb]{0,0.5,0} {-0.11} & \cellcolor[rgb]{0,0.5,0} {-0.08} & \cellcolor[rgb]{0,0.5,0} {-0.06} & \cellcolor[rgb]{0.351,0.035,0} {-0.02} & \cellcolor[rgb]{0.808,1,0} {0.02} & \cellcolor[rgb]{0,0.5,0} {0.07} & \cellcolor[rgb]{0,0.5,0} {0.06} & \cellcolor[rgb]{0,0.5,0} {0.07} & \cellcolor[rgb]{0,0.502,0} {0.04} & \cellcolor[rgb]{0,0.716,0} {0.03} & \cellcolor[rgb]{0.842,0.084,0} {-0.01} & \cellcolor[rgb]{0,0.5,0} {0} & \cellcolor[rgb]{0,0.5,0} {-0.01} & \cellcolor[rgb]{0.928,1,0} {0.02} & \cellcolor[rgb]{0,0,0} {0.01} & \cellcolor[rgb]{0,0.783,0} {0.03} & \cellcolor[rgb]{0,0,0} {-0.02} & \cellcolor[rgb]{0.198,1,0} {-0.03} & \cellcolor[rgb]{0,0.666,0} {-0.03} & \cellcolor[rgb]{0,0.5,0} {-0.07} \\ 
$\Delta P_{t-2}^{d,s}$ & \cellcolor[rgb]{0,0.5,0} {-0.05} & \cellcolor[rgb]{0,0.5,0} {-0.06} & \cellcolor[rgb]{0,0.504,0} {-0.04} & \cellcolor[rgb]{0,0.5,0} {-0.07} & \cellcolor[rgb]{0,0.501,0} {-0.05} & \cellcolor[rgb]{0,0.501,0} {-0.05} & \cellcolor[rgb]{0,0.532,0} {-0.04} & \cellcolor[rgb]{0,0.538,0} {-0.04} & \cellcolor[rgb]{0.609,1,0} {-0.02} & \cellcolor[rgb]{0,0,0} {0} & \cellcolor[rgb]{0,0,0} {0.01} & \cellcolor[rgb]{0.402,1,0} {0.02} & \cellcolor[rgb]{0,0,0} {0.01} & \cellcolor[rgb]{0,0,0} {0} & \cellcolor[rgb]{0,0.5,0} {0} & \cellcolor[rgb]{0,0.5,0} {0} & \cellcolor[rgb]{0,0.5,0} {-0.03} & \cellcolor[rgb]{0,0,0} {-0.01} & \cellcolor[rgb]{0,0,0} {0} & \cellcolor[rgb]{0,0.661,0} {-0.03} & \cellcolor[rgb]{0,0,0} {-0.01} & \cellcolor[rgb]{0,0,0} {-0.01} & \cellcolor[rgb]{0.223,0.022,0} {-0.02} & \cellcolor[rgb]{0,0.5,0} {-0.05} \\ 
$\Delta P_{t-3}^{d,s}$ & \cellcolor[rgb]{0,0,0} {-0.02} & \cellcolor[rgb]{0.007,1,0} {-0.03} & \cellcolor[rgb]{0,0.619,0} {-0.03} & \cellcolor[rgb]{0,0.805,0} {-0.03} & \cellcolor[rgb]{0,0.828,0} {-0.03} & \cellcolor[rgb]{0,0.505,0} {-0.04} & \cellcolor[rgb]{0,0,0} {-0.01} & \cellcolor[rgb]{0,0,0} {-0.01} & \cellcolor[rgb]{0,0,0} {0} & \cellcolor[rgb]{0,0,0} {0.01} & \cellcolor[rgb]{0.42,1,0} {0.02} & \cellcolor[rgb]{0,0,0} {0.01} & \cellcolor[rgb]{0,0,0} {0.01} & \cellcolor[rgb]{0,0,0} {0.01} & \cellcolor[rgb]{0,0,0} {0.01} & \cellcolor[rgb]{0,0.5,0} {0} & \cellcolor[rgb]{0,0.5,0} {0} & \cellcolor[rgb]{0,0,0} {0} & \cellcolor[rgb]{0,0,0} {0} & \cellcolor[rgb]{0,0,0} {-0.01} & \cellcolor[rgb]{0,0,0} {-0.01} & \cellcolor[rgb]{0,0,0} {0} & \cellcolor[rgb]{0,0,0} {0.01} & \cellcolor[rgb]{0,0,0} {-0.02} \\ 
\hline
	$g_2(\sigma_t^{d,s})$ & 0 & 1 & 2 & 3 & 4 & 5 & 6 & 7 & 8 & 9 & 10 & 11 & 12 & 13 & 14 & 15 & 16 & 17 & 18 & 19 & 20 & 21 & 22 & 23 \\ 
		\hline
		Intercept & \cellcolor[rgb]{0,0.5,0} {0.49} & \cellcolor[rgb]{0,0.5,0} {0.52} & \cellcolor[rgb]{0,0.5,0} {0.5} & \cellcolor[rgb]{0,0.5,0} {0.42} & \cellcolor[rgb]{0,0.5,0} {0.47} & \cellcolor[rgb]{0,0.5,0} {0.56} & \cellcolor[rgb]{0,0.5,0} {0.83} & \cellcolor[rgb]{0,0.5,0} {0.66} & \cellcolor[rgb]{0,0.5,0} {0.55} & \cellcolor[rgb]{0,0.5,0} {0.29} & \cellcolor[rgb]{0,0.5,0} {0.45} & \cellcolor[rgb]{0,0.5,0} {0.48} & \cellcolor[rgb]{0,0.5,0} {0.36} & \cellcolor[rgb]{0,0.5,0} {0.31} & \cellcolor[rgb]{0,0.5,0} {-1.78} & \cellcolor[rgb]{0,0.5,0} {-1.75} & \cellcolor[rgb]{0,0.5,0} {-1.65} & \cellcolor[rgb]{0,0.5,0} {0.39} & \cellcolor[rgb]{0,0.5,0} {0.62} & \cellcolor[rgb]{0,0.5,0} {0.36} & \cellcolor[rgb]{0,0.5,0} {0.36} & \cellcolor[rgb]{0,0.5,0} {0.59} & \cellcolor[rgb]{0,0.5,0} {0.61} & \cellcolor[rgb]{0,0.5,0} {0.51} \\ 
		$ \left|\Delta P_{t-1}^{d,s} \right|$ & \cellcolor[rgb]{0,0.5,0} {0.28} & \cellcolor[rgb]{0,0.5,0} {0.27} & \cellcolor[rgb]{0,0.5,0} {0.28} & \cellcolor[rgb]{0,0.5,0} {0.26} & \cellcolor[rgb]{0,0.5,0} {0.26} & \cellcolor[rgb]{0,0.5,0} {0.23} & \cellcolor[rgb]{0,0.5,0} {0.27} & \cellcolor[rgb]{0,0.5,0} {0.22} & \cellcolor[rgb]{0,0.5,0} {0.21} & \cellcolor[rgb]{0,0.5,0} {0.18} & \cellcolor[rgb]{0,0.5,0} {0.2} & \cellcolor[rgb]{0,0.5,0} {0.18} & \cellcolor[rgb]{0,0.5,0} {0.21} & \cellcolor[rgb]{0,0.5,0} {0.27} & \cellcolor[rgb]{0,0.5,0} {1.8} & \cellcolor[rgb]{0,0.5,0} {2.39} & \cellcolor[rgb]{0,0.5,0} {2.74} & \cellcolor[rgb]{0,0.5,0} {0.24} & \cellcolor[rgb]{0,0.5,0} {0.24} & \cellcolor[rgb]{0,0.5,0} {0.23} & \cellcolor[rgb]{0,0.5,0} {0.31} & \cellcolor[rgb]{0,0.5,0} {0.27} & \cellcolor[rgb]{0,0.5,0} {0.29} & \cellcolor[rgb]{0,0.5,0} {0.27} \\ 
		$ \left|\Delta P_{t-2}^{d,s} \right|$ & \cellcolor[rgb]{0,0.5,0} {0.09} & \cellcolor[rgb]{0,0.5,0} {0.13} & \cellcolor[rgb]{0,0.5,0} {0.12} & \cellcolor[rgb]{0,0.5,0} {0.15} & \cellcolor[rgb]{0,0.5,0} {0.16} & \cellcolor[rgb]{0,0.5,0} {0.16} & \cellcolor[rgb]{0,0.5,0} {0.14} & \cellcolor[rgb]{0,0.501,0} {0.08} & \cellcolor[rgb]{0,0.5,0} {0.1} & \cellcolor[rgb]{0,0.5,0} {0.07} & \cellcolor[rgb]{0,0.5,0} {0.1} & \cellcolor[rgb]{0,0.5,0} {0.1} & \cellcolor[rgb]{0,0.5,0} {0.12} & \cellcolor[rgb]{0,0.5,0} {0.14} & \cellcolor[rgb]{0,0.5,0} {-0.06} & \cellcolor[rgb]{0,0,0} {0} & \cellcolor[rgb]{0,0.5,0} {1.27} & \cellcolor[rgb]{0,0.5,0} {0.11} & \cellcolor[rgb]{0,0.5,0} {0.14} & \cellcolor[rgb]{0,0.5,0} {0.13} & \cellcolor[rgb]{0,0.5,0} {0.14} & \cellcolor[rgb]{0,0.5,0} {0.12} & \cellcolor[rgb]{0,0.5,0} {0.14} & \cellcolor[rgb]{0,0.5,0} {0.13} \\ 
		$ \left|\Delta P_{t-3}^{d,s} \right|$ & \cellcolor[rgb]{0,0.503,0} {0.07} & \cellcolor[rgb]{0,0.5,0} {0.08} & \cellcolor[rgb]{0,0.896,0} {0.05} & \cellcolor[rgb]{0,0.5,0} {0.08} & \cellcolor[rgb]{0,0.516,0} {0.06} & \cellcolor[rgb]{0,0.5,0} {0.08} & \cellcolor[rgb]{0,0.504,0} {0.06} & \cellcolor[rgb]{0,0.5,0} {0.08} & \cellcolor[rgb]{0,0.5,0} {0.08} & \cellcolor[rgb]{0,0.502,0} {0.05} & \cellcolor[rgb]{0,0.5,0} {0.06} & \cellcolor[rgb]{0,0.5,0} {0.05} & \cellcolor[rgb]{0,0.501,0} {0.05} & \cellcolor[rgb]{0,0.5,0} {0.09} & \cellcolor[rgb]{0,0.5,0} {1.1} & \cellcolor[rgb]{0,0.5,0} {-0.15} & \cellcolor[rgb]{0.696,1,0} {0.2} & \cellcolor[rgb]{0,0.5,0} {0.08} & \cellcolor[rgb]{0,0.513,0} {0.05} & \cellcolor[rgb]{0,0.585,0} {0.04} & \cellcolor[rgb]{0,0.502,0} {0.06} & \cellcolor[rgb]{0,0.5,0} {0.08} & \cellcolor[rgb]{0.559,1,0} {0.03} & \cellcolor[rgb]{0,0.502,0} {0.07} \\ 
		$ \left|\Delta P_{t-4}^{d,s} \right|$ & \cellcolor[rgb]{0,0,0} {0.03} & \cellcolor[rgb]{0,0,0} {0.02} & \cellcolor[rgb]{0,0.5,0} {0.09} & \cellcolor[rgb]{0.817,0.082,0} {0.03} & \cellcolor[rgb]{0,0.501,0} {0.07} & \cellcolor[rgb]{0,0.513,0} {0.06} & \cellcolor[rgb]{0.431,1,0} {0.03} & \cellcolor[rgb]{0,0.62,0} {0.04} & \cellcolor[rgb]{0,0.769,0} {0.03} & \cellcolor[rgb]{0,0.507,0} {0.04} & \cellcolor[rgb]{0,0.502,0} {0.05} & \cellcolor[rgb]{0,0.521,0} {0.04} & \cellcolor[rgb]{0,0.502,0} {0.05} & \cellcolor[rgb]{0,0.5,0} {0.07} & \cellcolor[rgb]{0.406,0.041,0} {0.02} & \cellcolor[rgb]{0,0,0} {0} & \cellcolor[rgb]{0,0.501,0} {-0.41} & \cellcolor[rgb]{0,0.505,0} {0.05} & \cellcolor[rgb]{0,0.503,0} {0.05} & \cellcolor[rgb]{0.979,0.098,0} {0.03} & \cellcolor[rgb]{0.403,1,0} {0.03} & \cellcolor[rgb]{0,0.503,0} {0.06} & \cellcolor[rgb]{0,0.507,0} {0.05} & \cellcolor[rgb]{0,0.873,0} {0.04} \\ 
		$ \left|\Delta P_{t-5}^{d,s} \right|$ & \cellcolor[rgb]{0.032,1,0} {0.04} & \cellcolor[rgb]{0,0,0} {0.02} & \cellcolor[rgb]{0,0,0} {0.02} & \cellcolor[rgb]{0,0.5,0} {0.07} & \cellcolor[rgb]{0.55,1,0} {0.03} & \cellcolor[rgb]{0,0.538,0} {0.05} & \cellcolor[rgb]{0,0,0} {0.02} & \cellcolor[rgb]{0.034,1,0} {0.04} & \cellcolor[rgb]{0,0.617,0} {0.04} & \cellcolor[rgb]{0.712,1,0} {0.02} & \cellcolor[rgb]{0,0.854,0} {0.03} & \cellcolor[rgb]{0,0.561,0} {0.04} & \cellcolor[rgb]{0.074,1,0} {0.03} & \cellcolor[rgb]{0,0.511,0} {0.05} & \cellcolor[rgb]{0,0.5,0} {-0.14} & \cellcolor[rgb]{0,0,0} {0} & \cellcolor[rgb]{0,0.501,0} {0.28} & \cellcolor[rgb]{0,0.666,0} {0.04} & \cellcolor[rgb]{0,0.788,0} {0.04} & \cellcolor[rgb]{0,0.811,0} {0.03} & \cellcolor[rgb]{0,0.514,0} {0.05} & \cellcolor[rgb]{0,0.608,0} {0.05} & \cellcolor[rgb]{0.904,0.09,0} {0.03} & \cellcolor[rgb]{0,0.505,0} {0.06} \\ 
		$ \left|\Delta P_{t-6}^{d,s} \right|$ & \cellcolor[rgb]{0.225,1,0} {0.04} & \cellcolor[rgb]{0.406,1,0} {0.04} & \cellcolor[rgb]{0.707,1,0} {0.04} & \cellcolor[rgb]{0.325,1,0} {0.04} & \cellcolor[rgb]{0.157,1,0} {0.04} & \cellcolor[rgb]{0.316,1,0} {0.04} & \cellcolor[rgb]{0,0.515,0} {0.05} & \cellcolor[rgb]{0.611,1,0} {0.03} & \cellcolor[rgb]{0.043,1,0} {0.03} & \cellcolor[rgb]{0,0.581,0} {0.03} & \cellcolor[rgb]{0,0.5,0} {0.06} & \cellcolor[rgb]{0,0.5,0} {0.06} & \cellcolor[rgb]{0,0.634,0} {0.03} & \cellcolor[rgb]{0,0.503,0} {0.05} & \cellcolor[rgb]{0,0,0} {0.02} & \cellcolor[rgb]{0,0,0} {0.01} & \cellcolor[rgb]{0,0,0} {0.02} & \cellcolor[rgb]{0,0.533,0} {0.04} & \cellcolor[rgb]{0.955,1,0} {0.02} & \cellcolor[rgb]{0,0.745,0} {0.04} & \cellcolor[rgb]{0.112,1,0} {0.04} & \cellcolor[rgb]{0,0.534,0} {0.05} & \cellcolor[rgb]{0.422,1,0} {0.03} & \cellcolor[rgb]{0,0.628,0} {0.05} \\ 
		$\sum_{j=7}^{12}  \left|\Delta P_{t-j}^{d,s} \right|$ & \cellcolor[rgb]{0,0.544,0} {0.06} & \cellcolor[rgb]{0,0.504,0} {0.08} & \cellcolor[rgb]{0,0.508,0} {0.07} & \cellcolor[rgb]{0,0.5,0} {0.09} & \cellcolor[rgb]{0,0.5,0} {0.14} & \cellcolor[rgb]{0,0.5,0} {0.09} & \cellcolor[rgb]{0,0.5,0} {0.09} & \cellcolor[rgb]{0,0.5,0} {0.1} & \cellcolor[rgb]{0,0.508,0} {0.05} & \cellcolor[rgb]{0,0.506,0} {0.05} & \cellcolor[rgb]{0,0.5,0} {0.06} & \cellcolor[rgb]{0,0.502,0} {0.05} & \cellcolor[rgb]{0,0.5,0} {0.1} & \cellcolor[rgb]{0,0.5,0} {0.08} & \cellcolor[rgb]{0,0,0} {-0.01} & \cellcolor[rgb]{0,0,0} {0} & \cellcolor[rgb]{0,0,0} {-0.01} & \cellcolor[rgb]{0,0.5,0} {0.08} & \cellcolor[rgb]{0,0.5,0} {0.08} & \cellcolor[rgb]{0,0.5,0} {0.1} & \cellcolor[rgb]{0,0.5,0} {0.11} & \cellcolor[rgb]{0,0.5,0} {0.15} & \cellcolor[rgb]{0,0.5,0} {0.1} & \cellcolor[rgb]{0,0.5,0} {0.13} \\ 
		$\text{DA}_{\text{Load}}^{d,s}$ & \cellcolor[rgb]{0,0,0} {-0.02} & \cellcolor[rgb]{0.489,1,0} {-0.03} & \cellcolor[rgb]{0,0,0} {-0.02} & \cellcolor[rgb]{0.997,1,0} {-0.03} & \cellcolor[rgb]{0,0,0} {0} & \cellcolor[rgb]{0,0,0} {0.02} & \cellcolor[rgb]{0.41,0.041,0} {0.05} & \cellcolor[rgb]{0,0,0} {-0.01} & \cellcolor[rgb]{0,0,0} {-0.02} & \cellcolor[rgb]{0,0,0} {-0.02} & \cellcolor[rgb]{0,0,0} {-0.01} & \cellcolor[rgb]{0,0,0} {0.02} & \cellcolor[rgb]{0.89,0.089,0} {0.03} & \cellcolor[rgb]{0,0,0} {-0.01} & \cellcolor[rgb]{0,0,0} {-0.03} & \cellcolor[rgb]{0.453,1,0} {-0.04} & \cellcolor[rgb]{0,0,0} {-0.02} & \cellcolor[rgb]{0,0,0} {-0.03} & \cellcolor[rgb]{0.962,0.096,0} {-0.04} & \cellcolor[rgb]{0,0,0} {0} & \cellcolor[rgb]{0,0,0} {-0.01} & \cellcolor[rgb]{0,0,0} {0} & \cellcolor[rgb]{0,0,0} {0.01} & \cellcolor[rgb]{0.326,1,0} {0.03} \\ 
		$\text{DA}_{\text{Sol}}^{d,s}$ & \cellcolor[rgb]{0,0,0} {0} & \cellcolor[rgb]{0,0,0} {0} & \cellcolor[rgb]{0,0,0} {0} & \cellcolor[rgb]{0,0,0} {0} & \cellcolor[rgb]{0.726,0.073,0} {-0.02} & \cellcolor[rgb]{0,0.643,0} {-0.04} & \cellcolor[rgb]{0.222,0.022,0} {-0.02} & \cellcolor[rgb]{0,0,0} {0.01} & \cellcolor[rgb]{0,0,0} {0.02} & \cellcolor[rgb]{0,0,0} {0.01} & \cellcolor[rgb]{0.352,0.035,0} {0.02} & \cellcolor[rgb]{0,0.542,0} {0.04} & \cellcolor[rgb]{0.043,1,0} {0.03} & \cellcolor[rgb]{0.315,0.031,0} {0.02} & \cellcolor[rgb]{0,0.75,0} {0.03} & \cellcolor[rgb]{0.115,1,0} {0.03} & \cellcolor[rgb]{0.974,0.097,0} {0.02} & \cellcolor[rgb]{0,0,0} {-0.01} & \cellcolor[rgb]{0,0.671,0} {-0.04} & \cellcolor[rgb]{0.589,1,0} {-0.03} & \cellcolor[rgb]{0,0.506,0} {-0.04} & \cellcolor[rgb]{0.104,1,0} {-0.03} & \cellcolor[rgb]{0,0.742,0} {-0.03} & \cellcolor[rgb]{0,0,0} {0} \\ 
		$\text{DA}_{\text{WiOn}}^{d,s}$ & \cellcolor[rgb]{0,0.5,0} {0.13} & \cellcolor[rgb]{0,0.5,0} {0.15} & \cellcolor[rgb]{0,0.5,0} {0.11} & \cellcolor[rgb]{0,0.5,0} {0.11} & \cellcolor[rgb]{0,0.5,0} {0.1} & \cellcolor[rgb]{0,0.532,0} {0.08} & \cellcolor[rgb]{0,0.501,0} {0.1} & \cellcolor[rgb]{0,0.501,0} {0.09} & \cellcolor[rgb]{0,0.5,0} {0.09} & \cellcolor[rgb]{0,0.5,0} {0.1} & \cellcolor[rgb]{0,0.5,0} {0.08} & \cellcolor[rgb]{0,0.5,0} {0.11} & \cellcolor[rgb]{0,0.5,0} {0.08} & \cellcolor[rgb]{0,0.5,0} {0.09} & \cellcolor[rgb]{0,0.5,0} {0.11} & \cellcolor[rgb]{0,0.5,0} {0.14} & \cellcolor[rgb]{0,0.5,0} {0.1} & \cellcolor[rgb]{0,0.5,0} {0.09} & \cellcolor[rgb]{0,0.5,0} {0.1} & \cellcolor[rgb]{0,0.5,0} {0.12} & \cellcolor[rgb]{0,0.5,0} {0.11} & \cellcolor[rgb]{0,0.5,0} {0.1} & \cellcolor[rgb]{0,0.5,0} {0.11} & \cellcolor[rgb]{0,0.5,0} {0.11} \\ 
		$\text{DA}_{\text{WiOff}}^{d,s}$ & \cellcolor[rgb]{0,0,0} {0.02} & \cellcolor[rgb]{0.591,0.059,0} {0.03} & \cellcolor[rgb]{0,0.502,0} {0.07} & \cellcolor[rgb]{0.393,1,0} {0.04} & \cellcolor[rgb]{0,0,0} {0.02} & \cellcolor[rgb]{0,0,0} {0.02} & \cellcolor[rgb]{0,0,0} {-0.01} & \cellcolor[rgb]{0,0,0} {0} & \cellcolor[rgb]{0,0,0} {0.01} & \cellcolor[rgb]{0.983,0.098,0} {0.03} & \cellcolor[rgb]{0.389,1,0} {0.03} & \cellcolor[rgb]{0.844,0.084,0} {0.03} & \cellcolor[rgb]{0.106,1,0} {0.03} & \cellcolor[rgb]{0,0,0} {0.01} & \cellcolor[rgb]{0,0,0} {0.02} & \cellcolor[rgb]{0,0,0} {0.02} & \cellcolor[rgb]{0.423,1,0} {0.03} & \cellcolor[rgb]{0,0,0} {-0.01} & \cellcolor[rgb]{0,0,0} {0} & \cellcolor[rgb]{0,0,0} {-0.01} & \cellcolor[rgb]{0,0,0} {0} & \cellcolor[rgb]{0,0,0} {0} & \cellcolor[rgb]{0,0,0} {0.01} & \cellcolor[rgb]{0,0,0} {0} \\ 
		$\text{Mon}(d)$ & \cellcolor[rgb]{0,0,0} {0.01} & \cellcolor[rgb]{0,0,0} {0.05} & \cellcolor[rgb]{0,0,0} {0.05} & \cellcolor[rgb]{0,0.752,0} {0.1} & \cellcolor[rgb]{0,0.551,0} {0.12} & \cellcolor[rgb]{0,0.5,0} {0.21} & \cellcolor[rgb]{0,0.507,0} {0.15} & \cellcolor[rgb]{0,0.747,0} {0.1} & \cellcolor[rgb]{0,0,0} {0.02} & \cellcolor[rgb]{0.824,1,0} {0.06} & \cellcolor[rgb]{0,0,0} {0.02} & \cellcolor[rgb]{0,0,0} {0.03} & \cellcolor[rgb]{0,0,0} {0.01} & \cellcolor[rgb]{0,0,0} {-0.01} & \cellcolor[rgb]{0,0,0} {0} & \cellcolor[rgb]{0,0,0} {-0.03} & \cellcolor[rgb]{0,0.795,0} {-0.08} & \cellcolor[rgb]{0,0.519,0} {-0.1} & \cellcolor[rgb]{0.539,0.054,0} {-0.05} & \cellcolor[rgb]{0,0,0} {-0.03} & \cellcolor[rgb]{0.56,1,0} {-0.06} & \cellcolor[rgb]{0.025,0.003,0} {-0.05} & \cellcolor[rgb]{0,0,0} {0.01} & \cellcolor[rgb]{0,0,0} {0} \\ 
		$\text{Sat}(d)$ & \cellcolor[rgb]{0,0,0} {-0.06} & \cellcolor[rgb]{0.623,0.062,0} {0.06} & \cellcolor[rgb]{0,0,0} {0.05} & \cellcolor[rgb]{0,0,0} {0.02} & \cellcolor[rgb]{0,0,0} {0.06} & \cellcolor[rgb]{0,0,0} {0.06} & \cellcolor[rgb]{0,0,0} {0.01} & \cellcolor[rgb]{0,0,0} {0.03} & \cellcolor[rgb]{0,0,0} {0.04} & \cellcolor[rgb]{0,0,0} {0.06} & \cellcolor[rgb]{0,0.504,0} {0.15} & \cellcolor[rgb]{0,0.5,0} {0.24} & \cellcolor[rgb]{0,0.5,0} {0.26} & \cellcolor[rgb]{0,0.528,0} {0.13} & \cellcolor[rgb]{0,0.504,0} {0.16} & \cellcolor[rgb]{0,0.5,0} {0.23} & \cellcolor[rgb]{0,0.5,0} {0.19} & \cellcolor[rgb]{0,0.507,0} {0.15} & \cellcolor[rgb]{0,0.615,0} {0.12} & \cellcolor[rgb]{0,0.501,0} {0.16} & \cellcolor[rgb]{0,0.688,0} {0.12} & \cellcolor[rgb]{0,0.963,0} {0.11} & \cellcolor[rgb]{0,0.99,0} {0.1} & \cellcolor[rgb]{0,0,0} {0.04} \\ 
		$\text{Sun}(d)$ & \cellcolor[rgb]{0.098,1,0} {0.11} & \cellcolor[rgb]{0,0,0} {0.07} & \cellcolor[rgb]{0,0.502,0} {0.18} & \cellcolor[rgb]{0.458,1,0} {0.09} & \cellcolor[rgb]{0.323,0.032,0} {0.08} & \cellcolor[rgb]{0,0,0} {0.06} & \cellcolor[rgb]{0,0,0} {-0.01} & \cellcolor[rgb]{0,0,0} {0.01} & \cellcolor[rgb]{0,0,0} {0.02} & \cellcolor[rgb]{0,0.881,0} {0.15} & \cellcolor[rgb]{0,0.571,0} {0.16} & \cellcolor[rgb]{0,0.5,0} {0.24} & \cellcolor[rgb]{0,0.5,0} {0.3} & \cellcolor[rgb]{0,0.503,0} {0.2} & \cellcolor[rgb]{0,0.502,0} {0.2} & \cellcolor[rgb]{0,0.5,0} {0.27} & \cellcolor[rgb]{0,0.5,0} {0.34} & \cellcolor[rgb]{0,0.501,0} {0.21} & \cellcolor[rgb]{0,0.5,0} {0.21} & \cellcolor[rgb]{0,0.5,0} {0.24} & \cellcolor[rgb]{0,0.5,0} {0.2} & \cellcolor[rgb]{0,0.816,0} {0.12} & \cellcolor[rgb]{0,0.542,0} {0.13} & \cellcolor[rgb]{0,0,0} {0.04} \\ 
		$\alpha_{t-1}^{d,s}$ & \cellcolor[rgb]{0,0.5,0} {-0.45} & \cellcolor[rgb]{0,0.5,0} {-0.44} & \cellcolor[rgb]{0,0.5,0} {-0.45} & \cellcolor[rgb]{0,0.5,0} {-0.39} & \cellcolor[rgb]{0,0.5,0} {-0.36} & \cellcolor[rgb]{0,0.5,0} {-0.39} & \cellcolor[rgb]{0,0.5,0} {-0.5} & \cellcolor[rgb]{0,0.5,0} {-0.4} & \cellcolor[rgb]{0,0.5,0} {-0.42} & \cellcolor[rgb]{0,0.5,0} {-0.38} & \cellcolor[rgb]{0,0.5,0} {-0.44} & \cellcolor[rgb]{0,0.5,0} {-0.35} & \cellcolor[rgb]{0,0.5,0} {-0.34} & \cellcolor[rgb]{0,0.5,0} {-0.37} & \cellcolor[rgb]{0,0.5,0} {-0.38} & \cellcolor[rgb]{0,0.5,0} {-0.34} & \cellcolor[rgb]{0,0.5,0} {-0.38} & \cellcolor[rgb]{0,0.5,0} {-0.4} & \cellcolor[rgb]{0,0.5,0} {-0.36} & \cellcolor[rgb]{0,0.5,0} {-0.32} & \cellcolor[rgb]{0,0.5,0} {-0.4} & \cellcolor[rgb]{0,0.5,0} {-0.4} & \cellcolor[rgb]{0,0.5,0} {-0.5} & \cellcolor[rgb]{0,0.5,0} {-0.39} \\ 
		$\alpha_{t-2}^{d,s}$ & \cellcolor[rgb]{0,0.5,0} {-0.18} & \cellcolor[rgb]{0,0.5,0} {-0.19} & \cellcolor[rgb]{0,0.5,0} {-0.18} & \cellcolor[rgb]{0,0.5,0} {-0.16} & \cellcolor[rgb]{0,0.5,0} {-0.17} & \cellcolor[rgb]{0,0.5,0} {-0.19} & \cellcolor[rgb]{0,0.5,0} {-0.22} & \cellcolor[rgb]{0,0.5,0} {-0.15} & \cellcolor[rgb]{0,0.5,0} {-0.15} & \cellcolor[rgb]{0,0.5,0} {-0.18} & \cellcolor[rgb]{0,0.5,0} {-0.15} & \cellcolor[rgb]{0,0.502,0} {-0.13} & \cellcolor[rgb]{0,0.5,0} {-0.15} & \cellcolor[rgb]{0,0.933,0} {-0.08} & \cellcolor[rgb]{0,0,0} {-0.04} & \cellcolor[rgb]{0,0,0} {-0.05} & \cellcolor[rgb]{0,0.5,0} {-0.17} & \cellcolor[rgb]{0,0.511,0} {-0.12} & \cellcolor[rgb]{0,0.5,0} {-0.2} & \cellcolor[rgb]{0,0.501,0} {-0.12} & \cellcolor[rgb]{0,0.501,0} {-0.13} & \cellcolor[rgb]{0,0.5,0} {-0.18} & \cellcolor[rgb]{0,0.5,0} {-0.16} & \cellcolor[rgb]{0,0.5,0} {-0.18} \\ 
		\hline
	
			$g_3(\nu_t^{d,s})$ & 0 & 1 & 2 & 3 & 4 & 5 & 6 & 7 & 8 & 9 & 10 & 11 & 12 & 13 & 14 & 15 & 16 & 17 & 18 & 19 & 20 & 21 & 22 & 23 \\ 
			\hline
		Intercept & \cellcolor[rgb]{0,0.5,0} {0.53} & \cellcolor[rgb]{0,0.5,0} {0.74} & \cellcolor[rgb]{0,0.5,0} {0.69} & \cellcolor[rgb]{0,0.5,0} {0.86} & \cellcolor[rgb]{0,0.5,0} {0.84} & \cellcolor[rgb]{0,0.5,0} {0.52} & \cellcolor[rgb]{0,0.5,0} {0.56} & \cellcolor[rgb]{0,0.5,0} {0.7} & \cellcolor[rgb]{0,0.5,0} {1.11} & \cellcolor[rgb]{0,0.5,0} {1.25} & \cellcolor[rgb]{0,0.5,0} {1.52} & \cellcolor[rgb]{0,0.5,0} {1.37} & \cellcolor[rgb]{0,0.5,0} {1.41} & \cellcolor[rgb]{0,0.5,0} {1.37} & \cellcolor[rgb]{0,0.5,0} {1.07} & \cellcolor[rgb]{0,0.5,0} {0.87} & \cellcolor[rgb]{0,0.5,0} {0.93} & \cellcolor[rgb]{0,0.5,0} {1.06} & \cellcolor[rgb]{0,0.5,0} {1.24} & \cellcolor[rgb]{0,0.5,0} {1.18} & \cellcolor[rgb]{0,0.5,0} {1.02} & \cellcolor[rgb]{0,0.5,0} {0.74} & \cellcolor[rgb]{0,0.5,0} {0.76} & \cellcolor[rgb]{0,0.5,0} {0.57} \\  
		\hline
	\end{tabular}
\end{adjustbox}
	\begingroup\tiny
	\setlength{\tabcolsep}{12pt}
	\begin{tabular}{rlllllllllll}
		& $ 0 \%$ & $ 1 \%$ & $ 2 \%$ & $ 3 \%$ & $ 4 \%$ & $ 5 \%$ & $ 6 \%$ & $ 7 \%$ & $ 8 \%$ & $ 9 \%$ & $ 10 \% \geq$ \\ 
		\hline
		Significance level & \cellcolor[rgb]{0,0.5,0} { } & \cellcolor[rgb]{0,1,0} { } & \cellcolor[rgb]{0.25,1,0} { } & \cellcolor[rgb]{0.5,1,0} { } & \cellcolor[rgb]{0.75,1,0} { } & \cellcolor[rgb]{1,1,0} { } & \cellcolor[rgb]{0.802,0.08,0} { } & \cellcolor[rgb]{0.601,0.06,0} { } & \cellcolor[rgb]{0.401,0.04,0} { } & \cellcolor[rgb]{0.2,0.02,0} { } & \cellcolor[rgb]{0,0,0} { } \\ 
		\hline
	\end{tabular}
	\endgroup
		\caption{Initial in-sample coefficient values of model \textbf{Mix.t.mu.sigma} reported for every hourly product. The price and generation variables were scaled for better clarity. The p-value of the test for significance of the values is indicated by the colour. The legend is explained by the table in the bottom.} 
		\label{tab:coefs}
	\end{table}

	The second sub-table shows coefficients of the model presented in equation \ref{eq:sigma_model}. Here, we see that all the variables using lagged absolute price differences are mostly significantly different from 0. Moreover, the coefficients of $|\Delta P_{t-1}^{d,s}|$ and $|\Delta P_{t-2}^{d,s}|$ are relatively high. Surprisingly, the day-ahead forecast of total load is mostly irrelevant. The day-ahead forecast of solar generation is significant mainly during the day-peak and in the evening. The day-ahead forecast of wind onshore generation appears to have a big positive impact on the volatility of price differences, in contrast to the wind offshore forecast. The behaviour of weekday dummies gives some light to our mixed expectancies -- they indicate a different behaviour of traders on Monday at night and on Saturday and Sunday during the day. On weekends the volatility is higher, likely due to higher bid-ask spreads on weekends. The lagged values of $\alpha_t^{d,s}$ have significant, negative impact on the volatility of price differences. It means that if there was trading at times $t-1$ and $t-2$, then the standard deviation would be lower. Let us note that the values of the coefficients are very similar among all hours except hours 14 to 16. For these hours the estimates of intercept and absolute price differences deviate heavily from the estimates of the remaining hours. A possible reason for this may be a few extreme outliers which were observed for these hours and for the others not.
	The table presents no values for the P-splines, because they are non-parametric functions. The last sub-table shows the estimate values for $g_3(\nu_t^{d,s})$ which we assumed to be constant. We show it anyway to gain an insight in the magnitude of the degrees of freedom. Let us recall that $g_3^{-1}(\nu) = \exp(\nu)+2$. Applying this to the estimate results in values of $\nu_t^{d,s}$ between around 3.7 and 6.6. Thus, the innovations are not extremely heavy tailed, and it is reasonable to apply asymptotic statistic for validation and interpretation.
	
			\begin{figure}[b!]
	\centering
	\includegraphics[width = \linewidth]{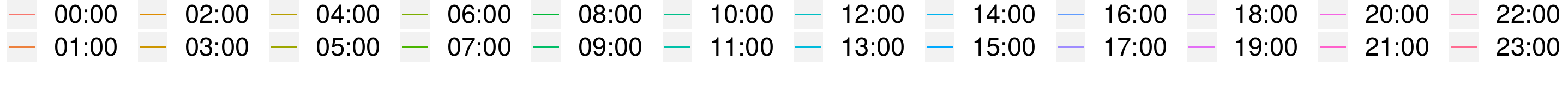}\\
	\vspace{-0.6cm}
	\subfloat[]{
		\includegraphics[width = 0.3165\linewidth]{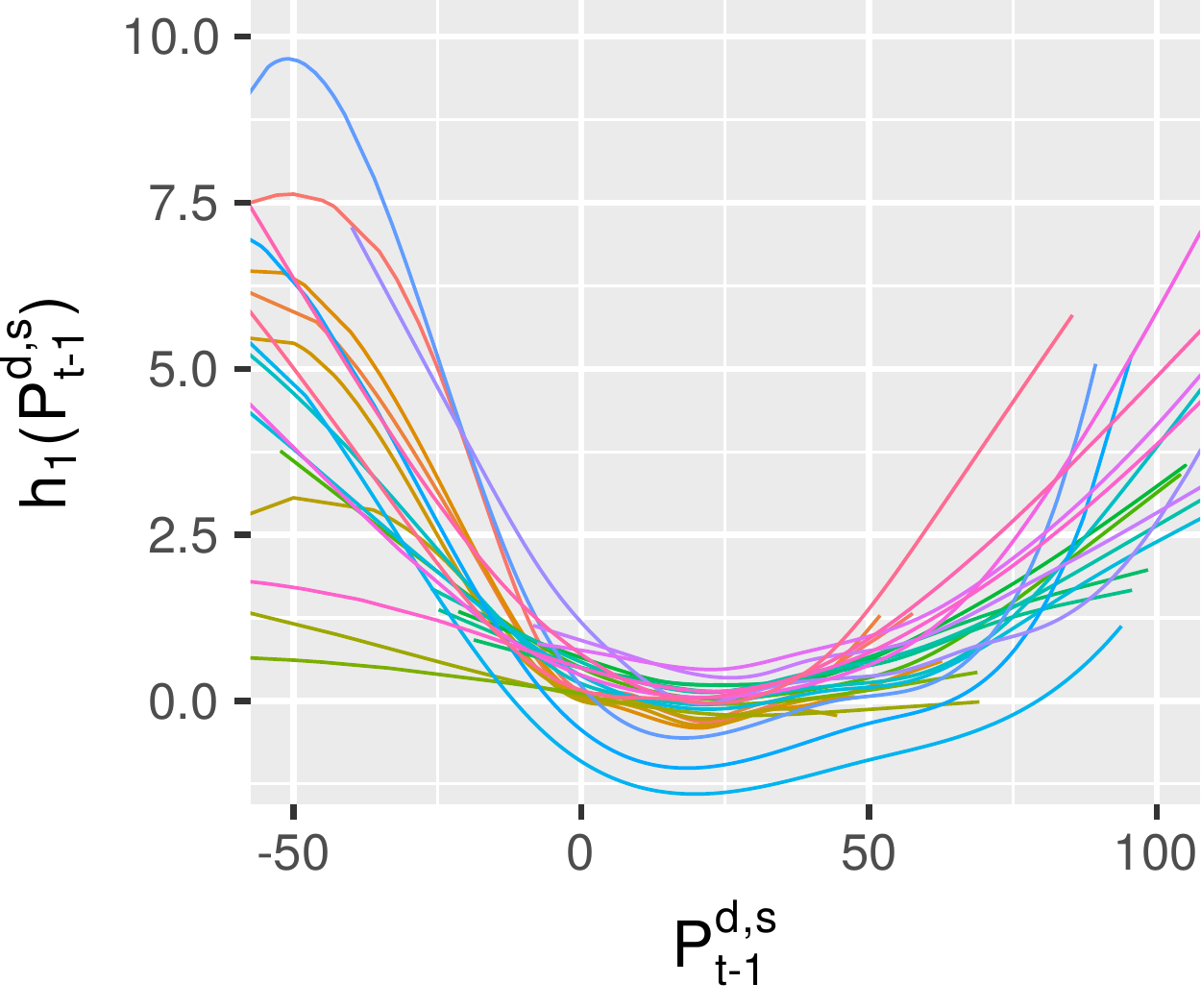}
		\label{fig:smoothing_price}
	}	
	\subfloat[]{
		\includegraphics[width = 0.3165\linewidth]{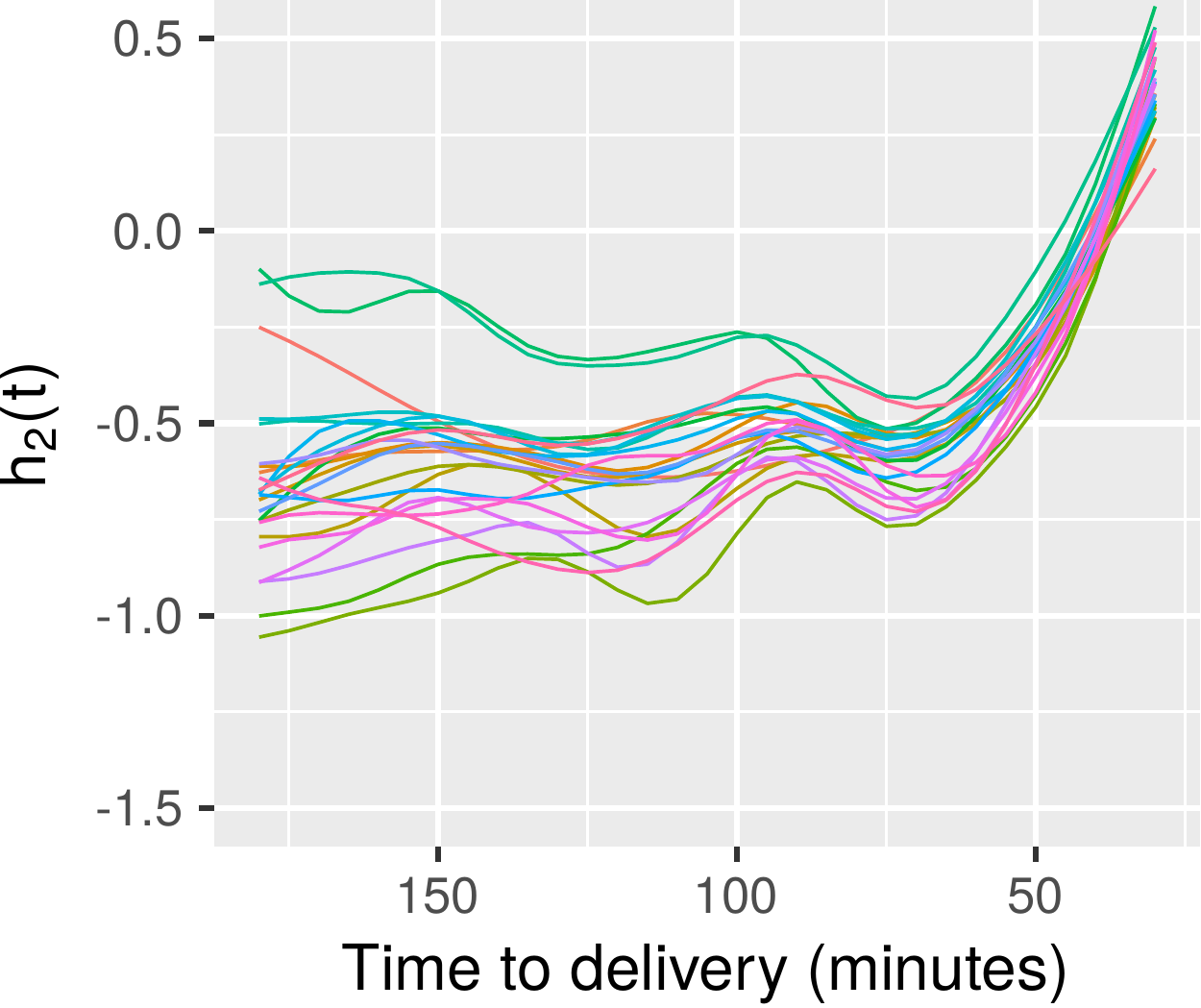}
		\label{fig:smoothing_ttm}
	}
	\subfloat[]{
		\includegraphics[width = 0.3165\linewidth]{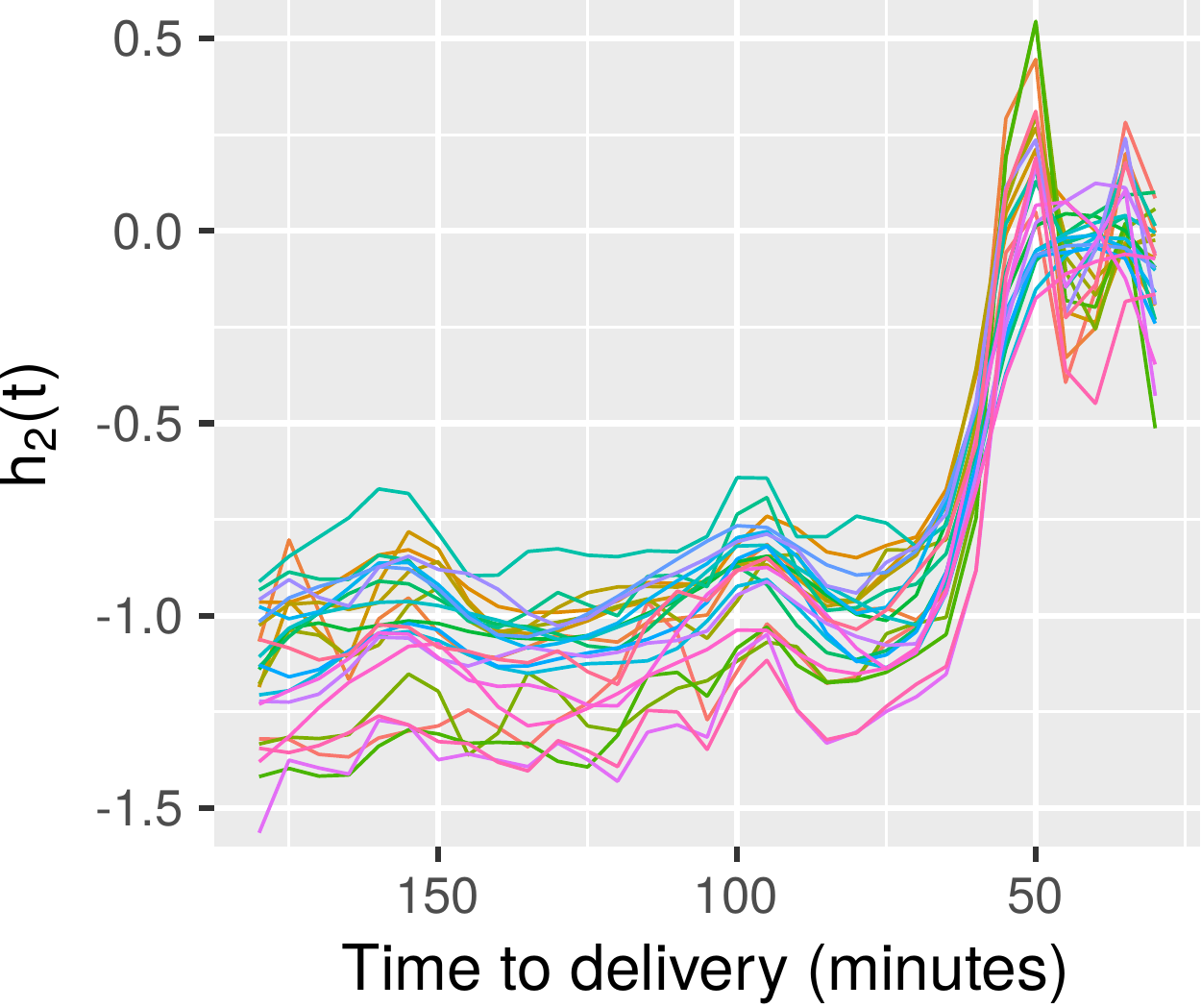}
		\label{fig:smoothing_ttm_xbid}
	}

	\caption{Initial in-sample smoothing effects of variables (a) $P_{t-1}^{d,s}$ and (b) time to maturity in the model described in equation \eqref{eq:sigma_model}. (c) is analogous to (b), but as in-sample considering the first year of XBID.  Note that the support of $P_{t-1}^{d,s}$ differs among products.}
	\label{fig:smoothing_effects}
\end{figure}
	
	Figure \ref{fig:smoothing_effects} shows the initial in-sample P-splines $h_1(P_{t-1}^{d,s})$ and $h_2(t)$. We see that in case of both variables, the smoothing functions are non-linear. Extreme values of most recent price $P_{t-1}^{d,s}$ result in most cases in high rise of volatility. On the other hand, the values between 0 and 50 EUR/MWh have rather marginal impact on the variance of the price differences. An interesting effect can be seen in Figure~\ref{fig:smoothing_ttm}. We see that until 60 minutes before the delivery the impact on the volatility is on a similar, negative level among all products. Then, in the last 30 minutes of trading the volatility rises substantially above zero. This behaviour can be misinterpreted as a result of the closure of XBID as in Figure~\ref{fig:market}. However, this plot is based on the initial in-sample data, i.e. the data between 16th July 2015 and 14th July 2016. Therefore, the effect of XBID could not be in the data as it was introduced on 18th June 2018. Figure~\ref{fig:smoothing_ttm_xbid} is analogous to Figure~\ref{fig:smoothing_ttm}, but based on the first year of XBID, i.e. the data between 18th June 2018 and 17th June 2019. Comparing the two figures concludes that the introduction of XBID has an impact on the volatility of the price differences decreasing it even lower before the XBID closure and rising it even higher just after it.
	 Interestingly, this is in contrary to the paper of \citet{kath2019modeling} who concluded that there is no evidence for the influence of XBID on the price volatility.

	\begin{figure}[b!]
		
		\includegraphics[width=1\textwidth]{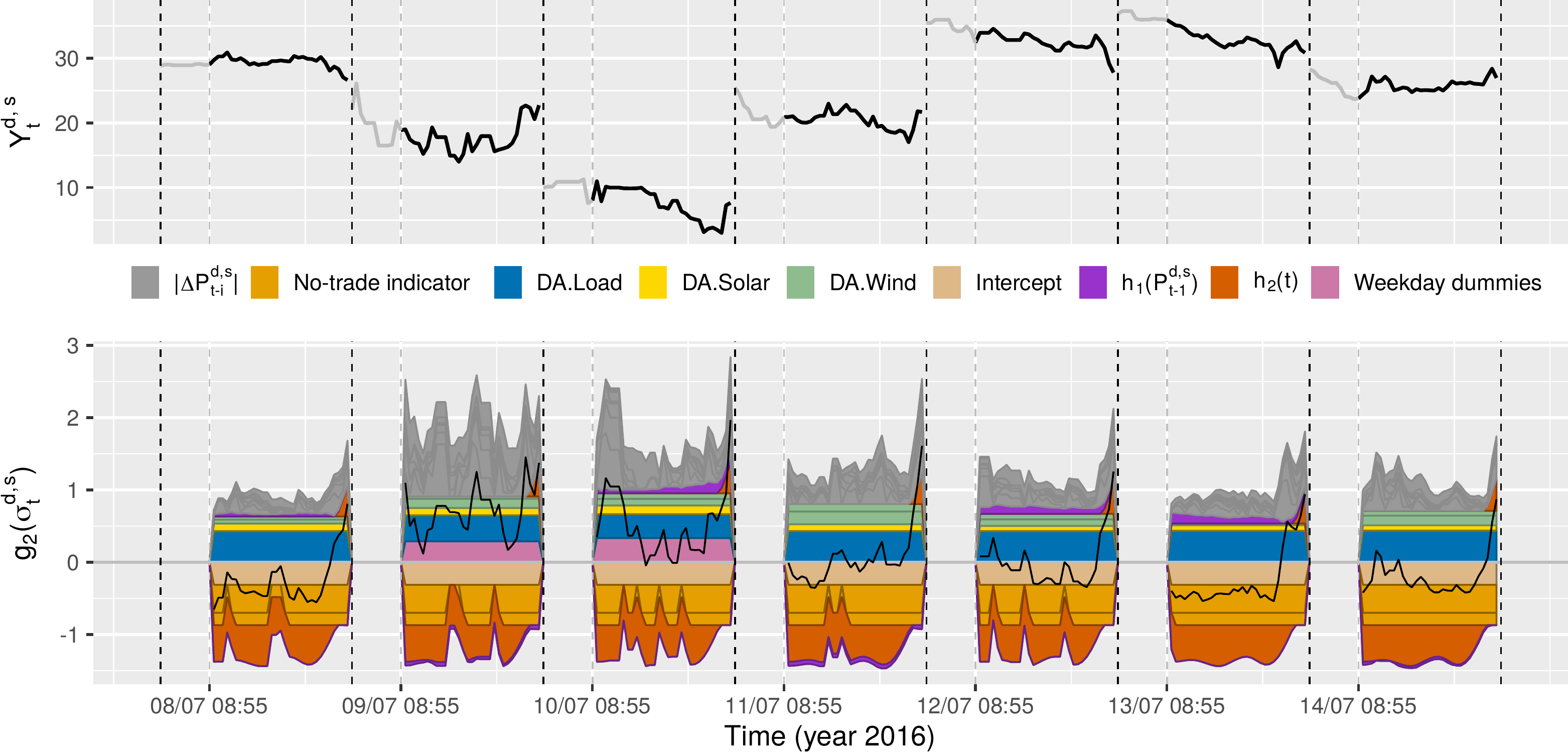}
		
		\caption{Price trajectory (top) and decomposition of fitted $g_2(\sigma_t^{d,s})$ (bottom) of hourly product with delivery at 12:00 for 7 consecutive in-sample days. The end of trading and the time of forecasting are indicated by the dashed black and grey lines, respectively.}
		\label{fig:sigma_comp}
		
	\end{figure}

	Figure~\ref{fig:sigma_comp} presents a price trajectory and a decomposition of fitted $g_2(\sigma_t^{d,s})$ of the product with delivery at 12:00 for the last 7 days of in-sample data.
% 	\FZC{Ich denke wir brauchen es nicht, wir wissen doch ob in-sample ein konstanter preis bleib bei positiven volumen oder bei keiner transaction.- ich würde es einfach löschen. For better clarity, in case of no-trade observation we assume no change in the decomposition. }
	For the sake of readability we grouped the components of the model for standard deviation similarly as in equation~\eqref{eq:sigma_model}. Let us note that the absolute price differences and fundamental regressors have big, positive impact on the volatility of price differences. 
	We also observe overall higher volatility on the weekend, i.e. the second and third day on the plot, than on the week. Note that 
	in this specific example the impact of the non-linear price due to $h_1$ looks rather negligible. However, the price level in these seven trading sessions is always between 0 and 40 EUR/MWh where we expect minor impacts.

	\subsection{Out-of-sample simulation}
\begin{figure}[b!]
		
		\includegraphics[width=1\textwidth]{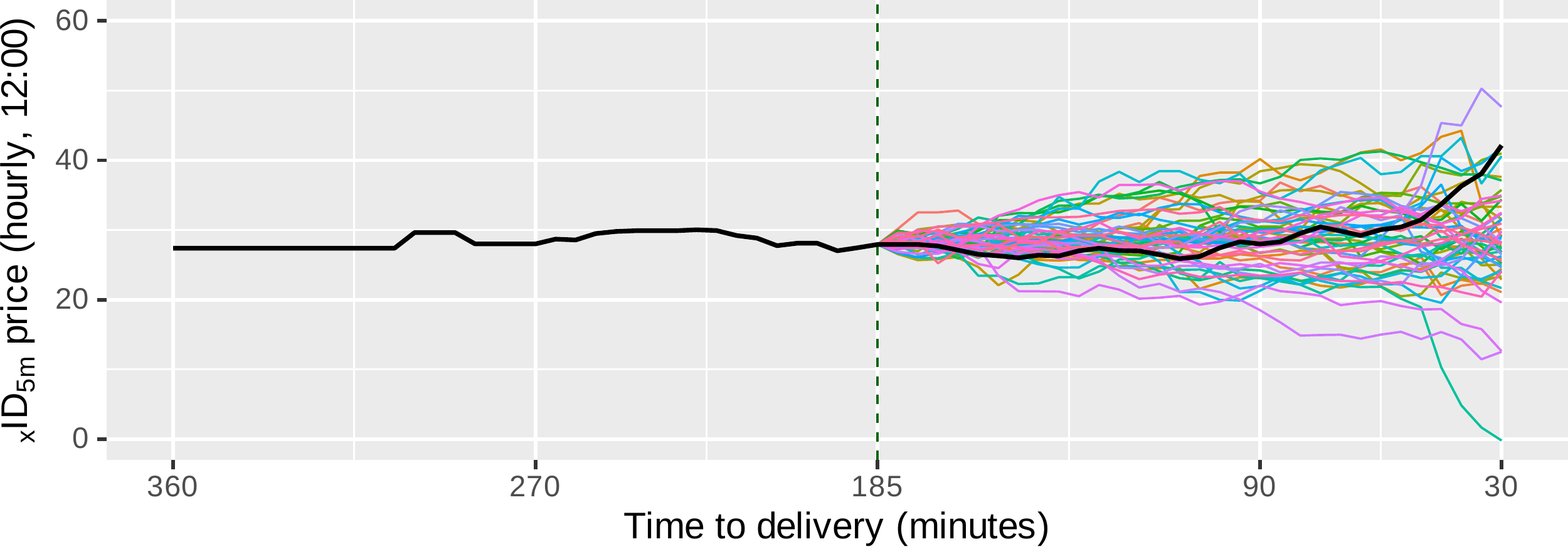}
		
		\caption{Price trajectory for the hourly product with delivery on 15.07.2016 at 12:00. The black part is the realization and the colourful part consists of 100 simulations from the \textbf{Mix.t.mu.sigma}. Time of forecasting is indicated by the green dashed line.}
		\label{fig:first_trajectory}
		
	\end{figure}

	Now, we turn ourselves to the analysis of the simulated trajectories. Figure~\ref{fig:first_trajectory} shows the first out-of-sample simulation exercise of prices of product with delivery at 12:00 on 15.07.2016. The trajectories are simulated from \textbf{Mix.t.mu.sigma} model and it can be easily compared to the simulations from Gaussian random walk presented in Figure~\ref{fig:motivation}. It is clear that in this example the trajectories of the mixture model are less volatile than the random walk.
	
\begin{table}[t!]
	\centering
	\begingroup\footnotesize\setlength{\tabcolsep}{5pt}
		%	\begin{adjustbox}{max width=1\textwidth}
\begin{tabular}{rrrrrrrrrr}
	\hline
	& ES & CRPS & VS & DSS & MAE & RMSE & 50\%-cov & 90\%-cov & 99\%-cov \\ 
	\hline
	Naive & \cellcolor[rgb]{1,0.996,0.5} {16.428} & \cellcolor[rgb]{0.953,1,0.5} {1.179} & \cellcolor[rgb]{0.875,1,0.5} {10.521} & \cellcolor[rgb]{1,0.5,0.55} {77.33} & \cellcolor[rgb]{0.612,0.937,0.5} {3.075} & \cellcolor[rgb]{0.553,0.918,0.5} {5.805} & \cellcolor[rgb]{0.629,0.943,0.5} {\textbf{0.490}} & \cellcolor[rgb]{0.834,1,0.5} {0.8919} & \cellcolor[rgb]{1,0.958,0.5} {0.9835} \\ 
	MV.N & \cellcolor[rgb]{1,0.846,0.5} {17.104} & \cellcolor[rgb]{1,0.892,0.5} {1.226} & \cellcolor[rgb]{1,0.885,0.5} {12.905} & \cellcolor[rgb]{1,0.5,0.528} {76.50} & \cellcolor[rgb]{1,0.982,0.5} {3.087} & \cellcolor[rgb]{0.793,0.998,0.5} {5.810} & \cellcolor[rgb]{1,0.729,0.5} {0.6712} & \cellcolor[rgb]{1,0.922,0.5} {0.9247} & \cellcolor[rgb]{1,0.716,0.5} {0.9723} \\ 
	MV.t & \cellcolor[rgb]{1,0.99,0.5} {16.454} & \cellcolor[rgb]{0.975,1,0.5} {1.181} & \cellcolor[rgb]{0.875,1,0.5} {10.524} & \cellcolor[rgb]{1,0.603,0.5} {71.66} & \cellcolor[rgb]{0.895,1,0.5} {3.081} & \cellcolor[rgb]{0.723,0.974,0.5} {5.809} & \cellcolor[rgb]{0.649,0.95,0.5} {0.5115} & \cellcolor[rgb]{0.727,0.976,0.5} {\textbf{0.8948}} & \cellcolor[rgb]{1,0.999,0.5} {0.9853} \\ 
	RW.N & \cellcolor[rgb]{1,0.5,0.55} {18.898} & \cellcolor[rgb]{1,0.5,0.55} {1.400} & \cellcolor[rgb]{1,0.5,0.55} {19.965} & \cellcolor[rgb]{1,0.5,0.513} {75.94} & \cellcolor[rgb]{1,0.866,0.5} {3.100} & \cellcolor[rgb]{0.936,1,0.5} {5.815} & \cellcolor[rgb]{1,0.5,0.55} {0.8025} & \cellcolor[rgb]{1,0.618,0.5} {0.9668} & \cellcolor[rgb]{0.665,0.955,0.5} {0.9887} \\ 
	RW.t & \cellcolor[rgb]{1,0.869,0.5} {16.999} & \cellcolor[rgb]{1,0.871,0.5} {1.234} & \cellcolor[rgb]{1,0.988,0.5} {11.218} & \cellcolor[rgb]{1,0.5,0.535} {76.77} & \cellcolor[rgb]{1,0.989,0.5} {3.086} & \cellcolor[rgb]{0.925,1,0.5} {5.815} & \cellcolor[rgb]{1,0.668,0.5} {0.6992} & \cellcolor[rgb]{1,0.683,0.5} {0.9578} & \cellcolor[rgb]{1,0.987,0.5} {0.9952} \\ 
	RW.t.mix.D & \cellcolor[rgb]{1,0.832,0.5} {17.168} & \cellcolor[rgb]{1,0.835,0.5} {1.248} & \cellcolor[rgb]{1,0.978,0.5} {11.385} & \cellcolor[rgb]{1,0.5,0.541} {77.01} & \cellcolor[rgb]{1,0.987,0.5} {3.086} & \cellcolor[rgb]{0.911,1,0.5} {5.814} & \cellcolor[rgb]{1,0.624,0.5} {0.7198} & \cellcolor[rgb]{1,0.656,0.5} {0.9615} & \cellcolor[rgb]{1,0.994,0.5} {0.9949} \\ 
	LQR.Gauss & \cellcolor[rgb]{1,0.972,0.5} {16.536} & \cellcolor[rgb]{1,0.993,0.5} {1.186} & \cellcolor[rgb]{0.691,0.964,0.5} { 9.923} & \cellcolor[rgb]{1,0.879,0.5} {61.36} & \cellcolor[rgb]{1,0.5,0.55} {3.162} & \cellcolor[rgb]{1,0.5,0.55} {5.966} & \cellcolor[rgb]{1,0.973,0.5} {0.5588} & \cellcolor[rgb]{1,0.953,0.5} {0.9204} & \cellcolor[rgb]{0.821,1,0.5} {0.9875} \\ 
	LQR.ind & \cellcolor[rgb]{1,0.959,0.5} {16.595} & \cellcolor[rgb]{1,0.98,0.5} {1.191} & \cellcolor[rgb]{0.822,1,0.5} {10.308} & \cellcolor[rgb]{1,0.959,0.5} {58.41} & \cellcolor[rgb]{1,0.5,0.55} {3.168} & \cellcolor[rgb]{1,0.5,0.55} {5.970} & \cellcolor[rgb]{1,0.929,0.5} {0.5789} & \cellcolor[rgb]{1,0.896,0.5} {0.9283} & \cellcolor[rgb]{0.71,0.97,0.5} {0.9884} \\ 
	Mix.RW.t & \cellcolor[rgb]{1,0.849,0.5} {17.092} & \cellcolor[rgb]{1,0.859,0.5} {1.239} & \cellcolor[rgb]{1,0.979,0.5} {11.377} & \cellcolor[rgb]{1,0.54,0.5} {73.97} & \cellcolor[rgb]{1,0.995,0.5} {3.085} & \cellcolor[rgb]{0.946,1,0.5} {5.815} & \cellcolor[rgb]{1,0.683,0.5} {0.6924} & \cellcolor[rgb]{1,0.689,0.5} {0.9569} & \cellcolor[rgb]{0.96,1,0.5} {0.9942} \\ 
	Mix.t.mu & \cellcolor[rgb]{1,0.807,0.5} {17.284} & \cellcolor[rgb]{1,0.818,0.5} {1.255} & \cellcolor[rgb]{1,0.962,0.5} {11.642} & \cellcolor[rgb]{1,0.526,0.5} {74.52} & \cellcolor[rgb]{1,0.984,0.5} {3.086} & \cellcolor[rgb]{0.951,1,0.5} {5.815} & \cellcolor[rgb]{1,0.641,0.5} {0.7117} & \cellcolor[rgb]{1,0.652,0.5} {0.9620} & \cellcolor[rgb]{1,0.991,0.5} {0.9950} \\ 
	Mix.t.sigma & \cellcolor[rgb]{0.513,0.904,0.5} {15.965} & \cellcolor[rgb]{0.508,0.903,0.5} {1.144} & \cellcolor[rgb]{0.514,0.905,0.5} { 9.444} & \cellcolor[rgb]{0.705,0.968,0.5} {54.42} & \cellcolor[rgb]{0.602,0.934,0.5} {3.075} & \cellcolor[rgb]{0.917,1,0.5} {5.814} & \cellcolor[rgb]{0.887,1,0.5} {0.5331} & \cellcolor[rgb]{1,0.98,0.5} {0.9167} & \cellcolor[rgb]{0.587,0.929,0.5} {\textbf{0.9907}} \\ 
	Mix.t.mu.sigma & \cellcolor[rgb]{0.5,0.9,0.5} {\textbf{15.956}} & \cellcolor[rgb]{0.5,0.9,0.5} {\textbf{1.144}} & \cellcolor[rgb]{0.5,0.9,0.5} {\textbf{9.405}} & \cellcolor[rgb]{0.5,0.9,0.5} {\textbf{53.15}} & \cellcolor[rgb]{0.5,0.9,0.5} {\textbf{3.073}} & \cellcolor[rgb]{0.5,0.9,0.5} {\textbf{5.804}} & \cellcolor[rgb]{1,0.996,0.5} {0.5482} & \cellcolor[rgb]{1,0.9,0.5} {0.9277} & \cellcolor[rgb]{0.839,1,0.5} {0.9928} \\ 
	\hline
\end{tabular}
%\end{adjustbox}
	\endgroup
	\caption{Error measures of the considered models. Colour indicates the performance columnwise (the greener, the better). With bold, we depicted the best values in each column.} 
	\label{tab:results}
\end{table}

Table~\ref{tab:results} shows the values of utilized error measures. The \textbf{Naive} model performs very well overall. Moreover, it gives the best results in terms of 50\% coverage. Very similar results to the naive model gives the \textbf{MV.t} which assumes the trajectories to follow a multivariate t-distribution. Having a look at the performance of \textbf{MV.N} we see that indeed the t-distribution is better in modelling of the trajectories. The LQR-based models are according to most of the considered measures worse than the \textbf{Naive} or \textbf{MV.t}. It is worth to emphasize the very bad ability to model the mean and median trajectories and on the other hand quite good 99\% coverage and the values of VS and DSS. Let us note a very bad performance of the Gaussian random walk. Model \textbf{RW.N} is clearly the worst. Having a look at its coverage values, we conclude that its simulations are too volatile. The introduction of t-distribution to random walk yields already a big improvement. Another step in our modelling, the usage of simple mixture distribution of the Dirac distribution~$\delta_0$ and the random walk with innovations from t-distribution do not improve the results. However, the next step, i.e. modelling of the probability $\pi_t^{d,s}$ with model \eqref{eq:logit_model} improves the results, but still they are not better than the ones of model \textbf{RW.t}. Moreover, modelling of the expected value as in equation \eqref{eq:mu_formula} also worsens the performance substantially. All these models are clearly worse than the \textbf{Naive} considering almost every measure. The last change to the mixture model, i.e. modelling of the standard deviation according to the formula in equation \eqref{eq:sigma_model} lowers the errors significantly. Modelling of the expected value in addition to the standard deviation brings a little improvement. Model \textbf{Mix.t.mu.sigma} is marginally better than \textbf{Mix.t.sigma} and it turns out to be the best model in terms of ES, CRPS,  VS, DSS, MAE and RMSE. A little disturbing are the values of the 50\%- and 90\%-coverage which are too high for the mixture models. This means that it is very likely that the results can be still improved. On the other hand, they capture better the behaviour in the tails than the \textbf{Naive} model.

\begin{table}[b!]
	\centering
	\begingroup\footnotesize\setlength{\tabcolsep}{5pt}
\begin{adjustbox}{max width=1\textwidth}
\begin{tabular}{rrrrrrrrrr}
	\hline
	& ES & CRPS & VS & DSS & MAE & RMSE & 50\%-cov & 90\%-cov & 99\%-cov \\ 
	\hline
	original & \cellcolor[rgb]{0.5,0.9,0.5} {15.956} & \cellcolor[rgb]{0.5,0.9,0.5} {1.144} & \cellcolor[rgb]{0.5,0.9,0.5} { 9.405} & \cellcolor[rgb]{0.5,0.9,0.5} {   53.15} & \cellcolor[rgb]{0.5,0.9,0.5} {3.073} & \cellcolor[rgb]{0.5,0.9,0.5} {5.804} & \cellcolor[rgb]{1,0.996,0.5} {0.5482} & \cellcolor[rgb]{1,0.9,0.5} {0.9277} & \cellcolor[rgb]{0.839,1,0.5} {0.9928} \\ 
	maximum dependency & \cellcolor[rgb]{1,0.5,0.55} {31.625} & \cellcolor[rgb]{0.5,0.9,0.5} {1.144} & \cellcolor[rgb]{0.814,1,0.5} {10.274} & \cellcolor[rgb]{1,0.5,0.55} { 9299.60} & \cellcolor[rgb]{0.5,0.9,0.5} {3.073} & \cellcolor[rgb]{0.5,0.9,0.5} {5.804} & \cellcolor[rgb]{1,0.996,0.5} {0.5482} & \cellcolor[rgb]{1,0.9,0.5} {0.9277} & \cellcolor[rgb]{0.839,1,0.5} {0.9928} \\ 
	minimum dependency & \cellcolor[rgb]{1,0.5,0.55} {33.263} & \cellcolor[rgb]{0.5,0.9,0.5} {1.144} & \cellcolor[rgb]{1,0.638,0.5} {16.918} & \cellcolor[rgb]{1,0.5,0.55} {27205.49} & \cellcolor[rgb]{0.5,0.9,0.5} {3.073} & \cellcolor[rgb]{0.5,0.9,0.5} {5.804} & \cellcolor[rgb]{1,0.996,0.5} {0.5482} & \cellcolor[rgb]{1,0.9,0.5} {0.9277} & \cellcolor[rgb]{0.839,1,0.5} {0.9928} \\ 
	independent & \cellcolor[rgb]{1,0.886,0.5} {16.925} & \cellcolor[rgb]{0.5,0.9,0.5} {1.144} & \cellcolor[rgb]{1,0.796,0.5} {14.350} & \cellcolor[rgb]{1,0.5,0.55} {  110.02} & \cellcolor[rgb]{0.5,0.9,0.5} {3.073} & \cellcolor[rgb]{0.5,0.9,0.5} {5.804} & \cellcolor[rgb]{1,0.996,0.5} {0.5482} & \cellcolor[rgb]{1,0.9,0.5} {0.9277} & \cellcolor[rgb]{0.839,1,0.5} {0.9928} \\ 
	\hline
\end{tabular}
\end{adjustbox}
	\endgroup
	\caption{Error measures of 4 \textbf{Mix.t.mu.sigma} models with different copulas.} 
	\label{tab:evidence}
\end{table}

The values of the error measures in Table~\ref{tab:results} may suggest that both ES and CRPS evaluate the same thing -- the marginal distribution. To emphasize that ES evaluates also the quality of the generated scenarios, we perform a small experiment on the model \textbf{Mix.t.mu.sigma}. In Table~\ref{tab:evidence}, we compare the performance of the \textbf{Mix.t.mu.sigma} with its copies modified using 3 copulas: maximum dependency, minimum dependency and independent. This results in the same marginal distributions, mean and median trajectories, and coverage values, but in completely different ensembles. This is depicted by the values of the measures -- the CRPS remains unchanged while the ES, VS and DSS changed drastically. Let us note the enormous aggravation of the DSS which is particularly sensitive to changes of the dependency structure.

\begin{figure}[b!]
	
	\includegraphics[width=1\textwidth]{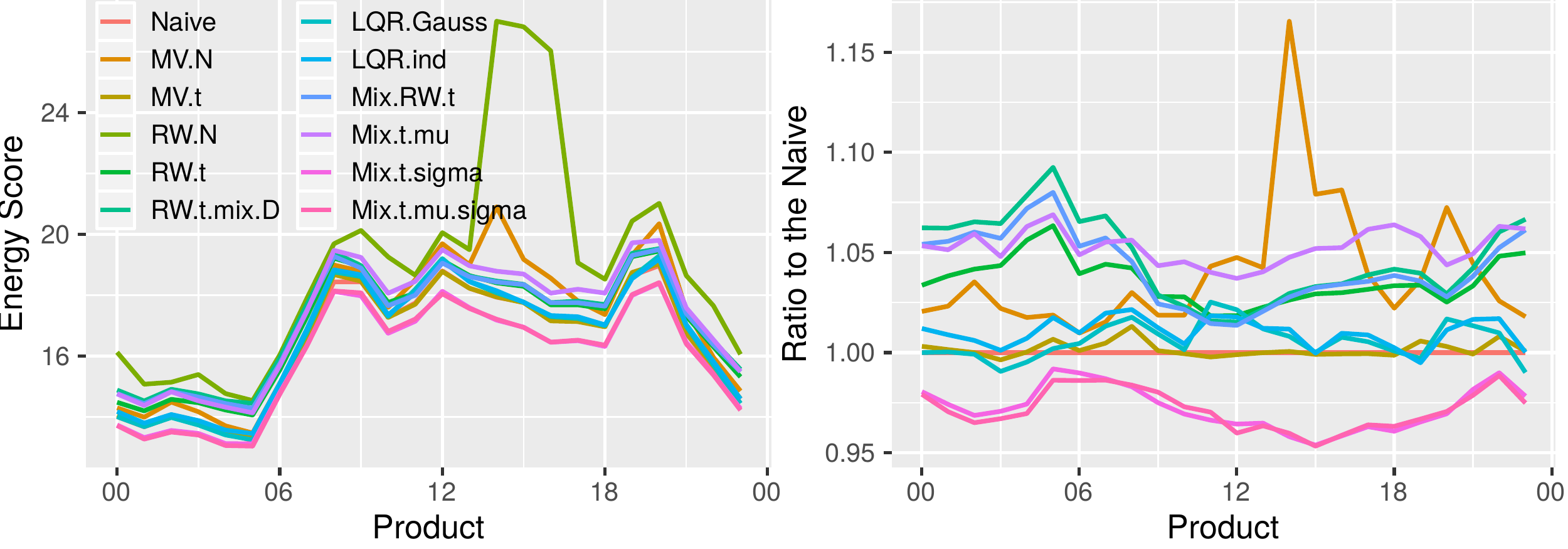}
	
	\caption{Energy score (left) and its ratio to the \textbf{Naive} (right) over 24 hourly products. The right graph is shown without \textbf{RW.N} for better clarity.}
	\label{fig:energy_score_plot}
	
\end{figure}

Figure~\ref{fig:energy_score_plot} shows the models' performance over all products in terms of energy score. A very interesting is the case of model \textbf{RW.N}. Usually it is not that much worse than the other random walks, but for hours 14-16 the error explodes. A look into the data explains the situation clearly -- there were a few in-sample observations of extreme price differences. The normal distribution assumes exponentially decaying tails, and thus the model overestimated the variance. This indicates clearly that the t-distribution is better in this purpose as it was unaffected by these events. Furthermore, we observe that models with modelled $\sigma_t^{d,s}$ are uniformly better than the others. 

\begin{figure}[b!]
	
	\includegraphics[width=1\textwidth]{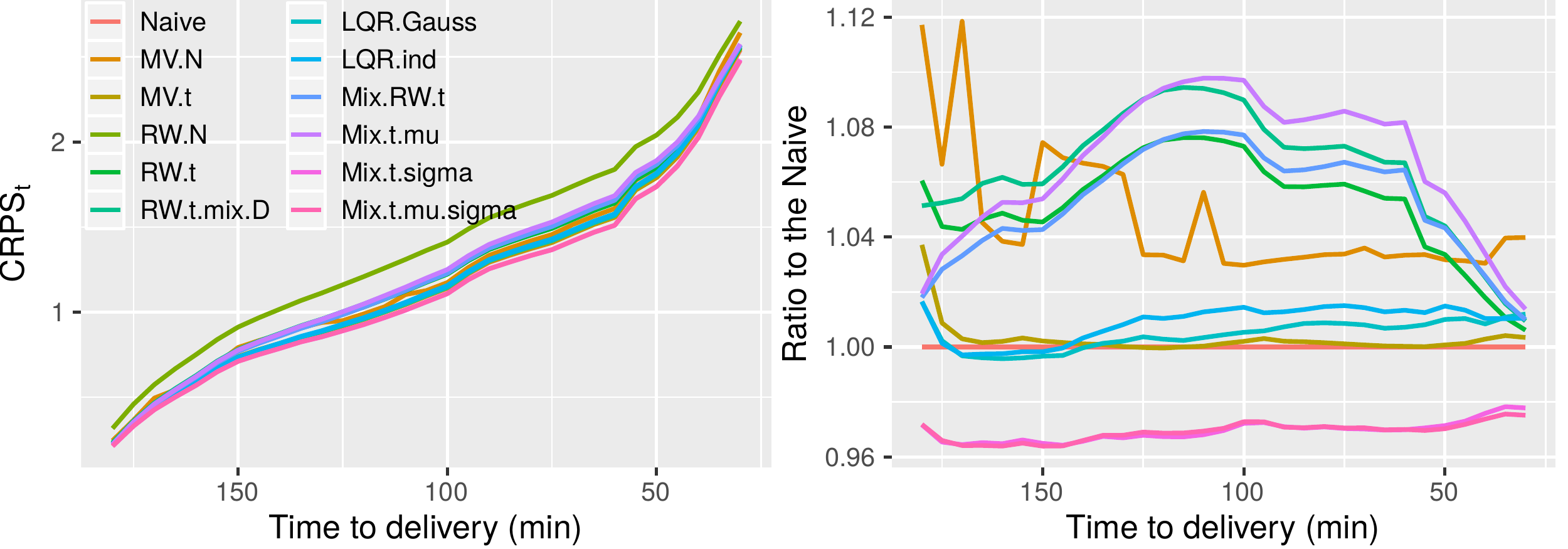}
	
	\caption{Continuous ranked probability score (left) and its ratio to the \textbf{Naive} (right) over time to delivery. The right graph is shown without \textbf{RW.N} for better clarity.}
	\label{fig:crps_over_ttm}
	
\end{figure}

Figure~\ref{fig:crps_over_ttm} presents the models' performance over time to delivery. The values rise as the time goes, but it is rather not surprising. An interesting behaviour can be observed from 150 to 100 minutes before the delivery. In this time range the errors of the random walk models and the mixture models that assume constant standard deviation rise significantly in comparison to the other models. It is also the time of decreasing volatility in Figure~\ref{fig:smoothing_ttm}.

			\begin{figure}[t!]
		
		\includegraphics[width=1\textwidth]{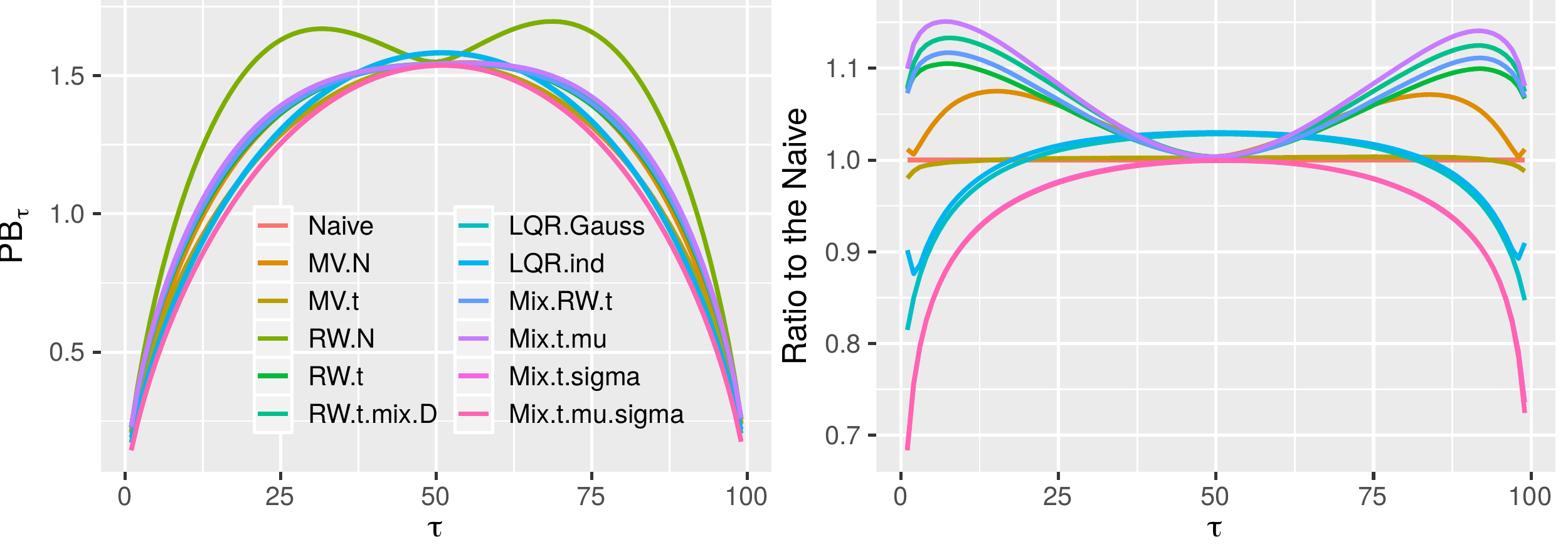}
		
		\caption{Pinball score (left) and its ratio to the \textbf{Naive} (right) over quantiles $\tau \in r$. The right graph is shown without \textbf{RW.N} for better clarity.}
		\label{fig:pinball_score_quants}
		
	\end{figure}
\begin{figure}[t!]
	\centering
	\subfloat[]{
		\includegraphics[width = 0.475\linewidth]{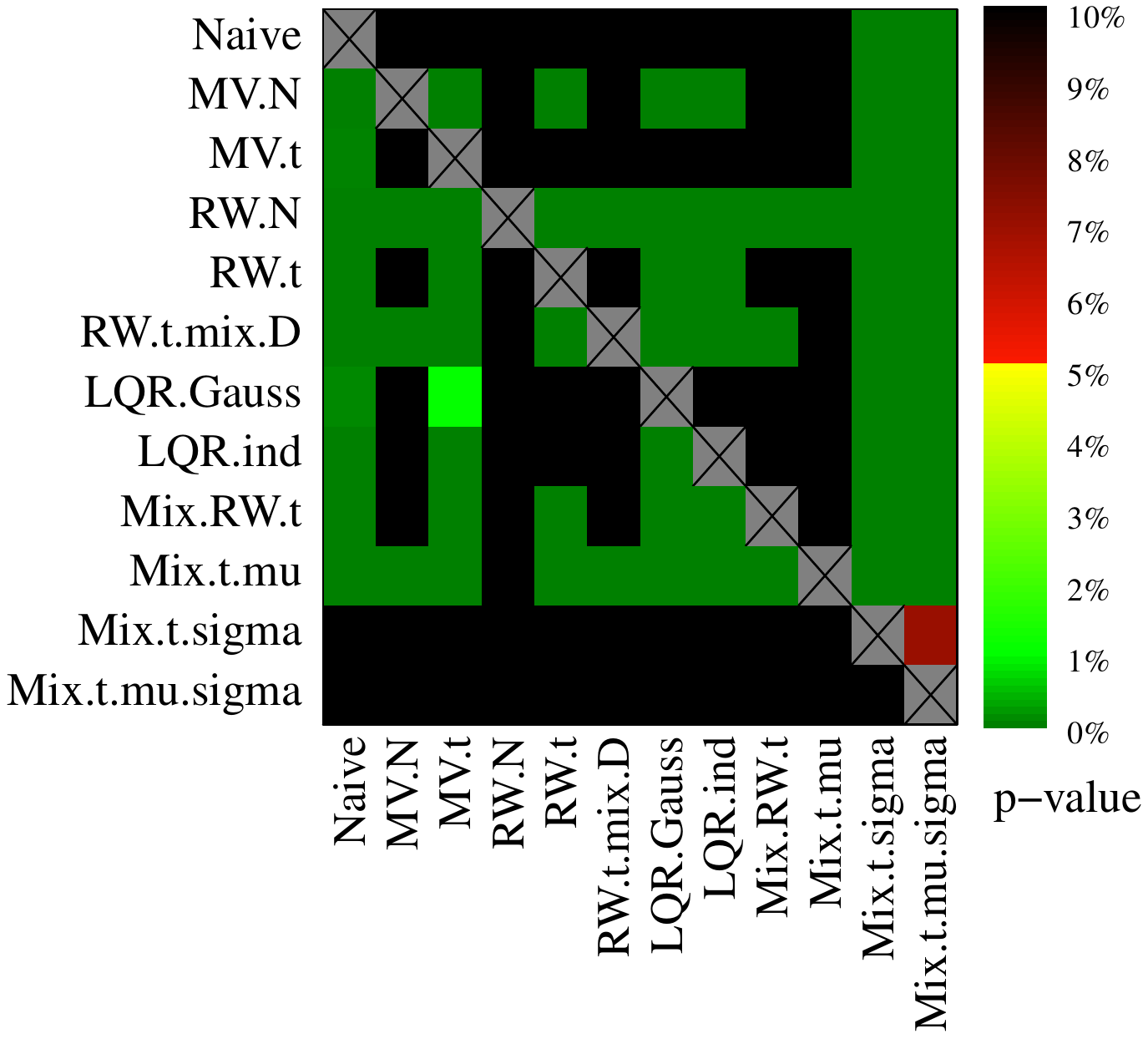}
		\label{fig:DM_ES}
	}	
	\subfloat[]{
		\includegraphics[width = 0.475\linewidth]{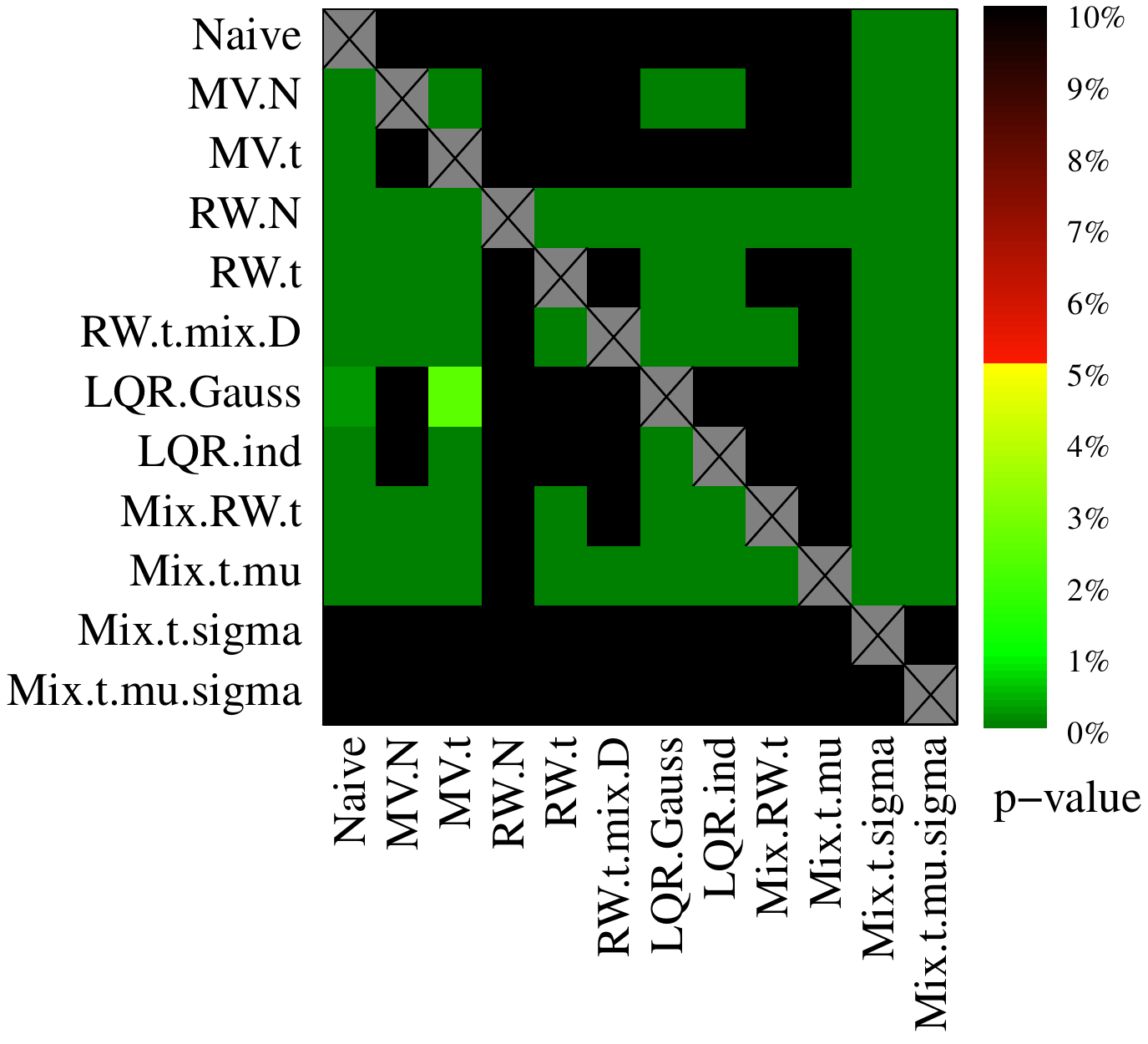}
		\label{fig:DM_CRPS}
	}
	\caption{Results of the Diebold-Mariano test. (a) presents the $p$-values for the $\text{ES}^{d,s}$ loss, (b) the values for the $\text{CRPS}^{d,s}$ loss. The figures use a heat map to indicate the range of the $p$-values. The closer they are to zero ($\to$ dark green), the more significant the difference is between forecasts of X-axis model (better) and forecasts of the Y-axis model (worse).}
	\label{fig:DMtest}
\end{figure}

Pinball Score values over quantiles $\tau$ are depicted in Figure~\ref{fig:pinball_score_quants}. Let us note that the gain from the forecasting of central quantiles is marginal, and it is inline with other studies regarding the ID$_3$-Price in the German intraday market \cite{narajewski2019econometric,janke2019forecasting}. On the other hand, models \textbf{Mix.t.sigma} and \textbf{Mix.t.mu.sigma} gain a lot from the forecasting of quantiles outside the centre, performing especially well in the tails. In relation to the naive benchmark, the error is around 30\% lower in the lower tail and around 25\% lower in the upper tail. Let us also note that the LQR-based models give quite good results in the tails, but lose a lot in the centre, compared to the naive or to the models with non-constant $\sigma_t^{d,s}$.

Figure \ref{fig:DMtest} shows the results of the DM test using two types of losses: the $\text{ES}^{d,s}$ and the $\text{CRPS}^{d,s}$. Before applying the test we conducted on the multivariate loss differential series $\Delta_{A,B}^d$ three tests that evaluate  the null hypothesis that a unit root is present in the series against the alternative that the data is stationary or trend-stationary. We used the Dickey-Fuller test \cite{dickey1979distribution}, the Augmented Dickey-Fuller test \cite{said1984testing}, and the Phillips-Perron test \cite{phillips1988testing}. In 99\% of cases the obtained p-values were smaller than 0.01, so we reject the null hypothesis. This indicates in our case that the loss differential series is covariance stationary. Only for the differences with \textbf{RW.N} the Dickey-Fuller test reported no significance for rejecting the null hypothesis. This may be caused by the bad capturing of the marginal distribution of the \textbf{RW.N}. The results of the DM test show that the difference between the forecasts of models \textbf{Mix.t.mu.sigma} and \textbf{Mix.t.sigma} is insignificant. Moreover, these models give better forecasts than all the other considered models. It is worth to emphasize a very good performance of the \textbf{Naive} model, but it is not surprising after taking a look at Table~\ref{tab:results}.

%\subsection{Trading application example}
%	

	\section{Conclusion}
	
	We conducted an ensemble forecasting study in the German Intraday Continuous Market which is novel in two ways. The first way, this study is the first one that raises the issue of price trajectory simulation and ensemble forecasting in continuous intraday electricity markets. The second way, the study uses a very clever mixture of distributions that is fitted to the data. The results are very satisfying and showing that it is possible to successfully model the volatility in the German Intraday Continuous Market. The study was carried out using the data from the German market, but the generality of this method and the organization of the European electricity markets ensure a possible application to other markets, especially the markets participate in XBID which covers exchanges like EPEX, Nordpool and OMIE.
	
	Obviously, the proposed method can be developed further. One of possible directions is using other external processes like the traded volume or price of nearby hours as regressors. Although, this is a non-trivial task and could easily lead to the accumulation of errors. Another possibility is utilization of other probability distribution. The not perfect coverage of the best performing model indicates that there is still some space for improvement. This issue could be addressed with some post-processing method as well.
	
	\section*{Acknowledgments}
	
	This research article was partially supported by the German Research Foundation (DFG, Germany) and the National Science Center (NCN, Poland) through BEETHOVEN grant no. 2016/23/G/HS4/01005.
	
	%\nocite{*}
	\vspace{-5mm} 
	\bibliographystyle{unsrtnat1}

	\bibliography{bibliography}	

\end{document}